\documentclass[aps,prc,showpacs,superscriptaddress,showkeys,twocolumn]{revtex4}
\usepackage{latexsym}
\usepackage{amssymb}
\usepackage{amsmath}
\usepackage{graphicx}
\usepackage{bm}
\usepackage{epsf}
\usepackage{epstopdf}

\newcommand{\nc}{\newcommand}       
\nc{\beq}       {\begin{eqnarray}}
\nc{\eeq}       {\end{eqnarray}}
\nc{\nn}        {\nonumber\\}
\nc{\mb}        {\mathbf}

\begin{document}

\title{Effects of nuclear symmetry energy and equation of state on neutron star properties}
\author{Fan Ji}
\affiliation{School of Physics, Nankai University, Tianjin 300071, China}
\author{Jinniu Hu}~\email{hujinniu@nankai.edu.cn}
\affiliation{School of Physics, Nankai University, Tianjin 300071, China}
\author{Shishao Bao}~\email{bao_shishao@163.com}
\affiliation{School of Physics and Information Engineering, Shanxi Normal University, Linfen 041004, China}
\author{Hong Shen}~\email{shennankai@gmail.com}
\affiliation{School of Physics, Nankai University, Tianjin 300071, China}

\begin{abstract}
We study the effects of nuclear symmetry energy on the mass-radius relation and
tidal deformability of neutron stars, considering the self-consistency of the equation
of state (EOS). We first construct a set of unified EOSs based on
relativistic mean-field models with different density
dependence of the symmetry energy.
For the description of pasta phases appearing in the inner crust of neutron stars,
we perform a self-consistent Thomas--Fermi calculation using the same nuclear interaction
as that for the uniform matter in the core. To examine possible effects from
the self-consistency of the EOS on neutron-star properties, we separately investigate
the impacts of crust and core segments. By matching the same core EOS to different crust EOSs,
it is found that neutron-star radii are significantly affected by the crust segment.
On the other hand, the neutron-star radii are also strongly dependent on the core EOS.
However, the correlation between the radius and the symmetry energy slope of the core EOS is
opposite to that of the crust EOS. It is likely that the nuclear model
with a small slope parameter is favored by recent astrophysical observations.
\end{abstract}

\maketitle


\section{Introduction}
\label{sec:1}

The recent detection of gravitational waves from a binary neutron-star
merger, known as GW170817, provides an upper limit on the tidal deformability
of neutron stars~\cite{Abbo17,Abbo18}, which can be used to constrain the equation
of state (EOS) of dense matter.
Another strong constraint on the EOS comes from the observations of massive
pulsars, PSR J1614-2230~\cite{Demo10,Fons16}, PSR J0348+0432~\cite{Anto13},
and PSR J0740+6620~\cite{Crom19},
which requires the maximum neutron-star mass to be larger than $\approx 2 M_\odot$.
The EOS plays a decisive role in understanding various properties of neutron
stars~\cite{Latt04,Cham08,Latt16,Oert17}.
Generally, the EOS used for the calculations of neutron-star structure must cover
a wide density range that can be divided into three segments:
(a) the EOS of the outer crust below the neutron drip density;
(b) the EOS of the inner crust from neutron drip to crust-core transition;
(c) the EOS of the liquid core above the crust-core transition.
The outer crust, which extends from the surface of the star to the neutron drip density,
is composed of spherical nuclei arranged in a lattice and a background of relativistic
electron gas. The behavior of the outer crust is mainly determined by experimental
nuclear masses, and therefore no significant differences exist in the EOS of the outer
crust when using different nuclear-mass models~\cite{Pear11}.
With increasing depth in the neutron star, neutrons drip out of nuclei and form a dilute
neutron gas together with the electron gas in the inner crust.
As the density increases toward the crust-core transition, spherical nuclei may become unstable
and nuclear shape is likely to change from droplet to rod, slab, tube, and bubble,
known as nuclear pasta phases~\cite{Rave83,Mene08,Bao14a}.
The inner crust of neutron stars has received much attention due to its
complex structure and important role in astrophysical
observations~\cite{Gril12,Okma13,Bao15,Fatt17}.
The transition from the inner crust to the core occurs at about $1/3$ to $1/2$ nuclear
saturation density depending on the nuclear interaction used.
The uniform matter in the core consists of a mixture of neutrons, protons, electrons,
and muons in $\beta$ equilibrium, which extends from the crust-core transition to
the center of the star. In the deep interior of neutron stars, non-nucleonic degrees
of freedom, such as hyperons and quarks, may appear to soften the EOS, as extensively
discussed in the literature~\cite{Oert17,Glen92,Glen01,Webe05,Masu13}.
In the inner core region of massive neutron stars, the baryon density can reach
higher than five times nuclear saturation density, where the deconfinement
hadron-quark phase transition probably occurs.
The phase transition from hadronic matter to quark matter is often
assumed to proceed with the Gibbs or Maxwell constructions,
depending on the surface tension at the hadron-quark interface~\cite{Bhat10}.
In addition, the hadron-quark pasta phases could be formed as a result of the
competition between the Coulomb and surface energies, as discussed
in Refs.~\cite{Glen01,Yasu14,Webe16,Wu19}.
The appearance of a hadron-quark phase transition would lead to a reduction of
the maximum neutron-star mass, but it is still possible to be compatible with
the $\approx 2 M_\odot$ constraint~\cite{Yasu14,Webe16,Wu19}.
For simplicity, we do not include non-nucleonic degrees of freedom in the present study.

It is important to investigate neutron-star properties using the unified EOS,
where both the core and the crust are based on the same nuclear interaction model.
There are several works~\cite{Haen01,Shen02,Miya13,Cham13,Shar15,Fort16} on developing the
unified EOS. A compressible liquid drop model was used to
describe the nuclei in the crust in Ref.~\cite{Haen01}, whereas the parameterized
Thomas--Fermi method~\cite{Shen02,Miya13} and self-consistent Thomas--Fermi
approximation~\cite{Cham13,Shar15,Fort16} were employed for nonuniform matter in
the crust region.
The use of a unified EOS is important for the description of the crust-core transition and
detailed properties of neutron stars.
However, in most calculations, a nonunified EOS is employed, i.e., the core EOS is matched
to a crust EOS obtained from different models. It was found in Ref.~\cite{Fort16} that
the matching procedure could slightly affect the resulting radius and crust thickness
of neutron stars. It is often argued that the crust EOS has less effect on the
global properties of neutron stars.
Considering recent observational progress related to neutron stars, we would like to
quantitatively examine the effect of matching different crust EOSs to the core EOS
on neutron-star properties such as the radius and tidal deformability.
For this purpose, we construct a set of EOSs for both the inner crust and the core,
employing the relativistic mean-field (RMF) model~\cite{Sero86,Meng06,TM1}.

It is well known that the nuclear symmetry energy and its density dependence
play an important role in understanding many phenomena in nuclear physics
and astrophysics~\cite{Latt16,Oert17,LiBA08,Horo01,Duco10}.
It has been found that various properties of neutron stars, such as the radius and
the crust structure, are sensitive to the symmetry energy $E_{\rm sym}$
and its slope parameter $L$~\cite{Latt16,Gril12,Bao15,Oyam07,Mene11,Prov13}.
In recent decades, great efforts have been devoted to constraining the values
of $E_{\rm sym}$ and $L$ at saturation density based on astrophysical
observations and terrestrial nuclear
experiments~\cite{Tews17,Hebe13,Latt14,Hage15,Dani14,Roca15,Birk17,Dani17}.
In Ref.~\cite{Oert17}, a sufficient number of constraints on the symmetry
energy parameters have been summarized, and the most probable values
for the symmetry energy and its slope at saturation density
were found to be $E_{\rm sym}=31.7\pm 3.2$ MeV and $L=58.7\pm 28.1$ MeV, respectively,
with a much larger error for $L$ than that for $E_{\rm sym}$.
To study the effect of the symmetry energy on neutron-star properties,
we employ a set of generated RMF models based on the TM1 parametrization,
which was described in our previous work~\cite{Bao14b}.
The original TM1 model~\cite{TM1} could provide satisfactory description for finite nuclei,
and meanwhile it has also been successfully used to construct the EOS
for supernova simulations and neutron stars~\cite{Oert17,Shen02,Shen11}.
We introduce an additional $\omega$-$\rho$ coupling term in the TM1 model,
which plays an essential role in controlling the density dependence of the symmetry
energy~\cite{Mene11,Prov13,Bao14b,IUFSU}.
By adjusting simultaneously two parameters associated to the $\rho$ meson
($g_{\rho}$ and ${\Lambda}_{\rm{v}}$), we can generate a model with a given $L$
at saturation density and a fixed $E_{\rm{sym}}$ at a density of 0.11 fm$^{-3}$.
The choice of fixing symmetry energy at 0.11 fm$^{-3}$ is
based on the following consideration. The generated models with different $L$
should be able to provide results for finite nuclei similar to the original TM1 model.
It is well known that the binding energy of finite nuclei is essentially determined
by the symmetry energy at $\approx 0.11$ fm$^{-3}$, not by the symmetry energy
at saturation density. Therefore, the value of the symmetry energy
at $\approx 0.11$ fm$^{-3}$ is well constrained by experimental nuclear masses.
By keeping $E_{\rm{sym}}$ fixed at $n_b=0.11\, \rm{fm}^{-3}$,
the resulting binding energies of finite nuclei are almost unchanged within
the set of generated models.
This is because the average value of the density in finite nuclei is about 0.11 fm$^{-3}$.
It is noteworthy that all models in this set have the same isoscalar
properties and fixed symmetry energy at $n_b=0.11\, \rm{fm}^{-3}$,
but they have different density dependence of the symmetry energy.
Therefore, this set of models is suitable for studying the correlations between
the slope parameter $L$ and neutron-star properties.
In the present work, we use RMF models based on the TM1 parametrization
for the calculations of neutron-star properties, which may introduce
some model dependency in the results. Generally, the predicted properties
of neutron stars, such as gravitational masses and radii, are model dependent,
as can be found in Refs.~\cite{Latt16,Oert17,Fort16,Gand12}.
As an example, the radius of a canonical $1.4 M_\odot$ neutron star ($R_{1.4}$)
varies between $\approx 11$ and 15 km for some popular EOSs supporting the $\approx 2 M_\odot$
maximum mass constraint~\cite{Latt16,Oert17,Fort16,Gand12}.
Even for several models with a similar slope parameter $L$, the difference in $R_{1.4}$
can be as large as $\approx 1$ km. Therefore, the slope parameter $L$ cannot be
precisely constrained by observations of neutron-star radii.
However, a positive correlation between the slope parameter $L$ and neutron-star radius
is consistent among different models, which will be studied using the RMF models in
the present calculations.

We have two aims in this study. The first is to construct a set of
unified EOSs using the RMF models that have the same isoscalar
properties but different density dependence of the symmetry energy,
and then apply these EOSs to study the effects of the symmetry energy on
neutron-star properties.
The second is to examine separately the influences from the crust and
core segments on the radius and tidal deformability of neutron stars.
By matching different crust EOSs to a fixed core EOS, the uncertainty
induced by the crust segment in a nonunified EOS can be estimated
quantitatively.
For constructing a unified EOS, we perform the self-consistent Thomas--Fermi
calculations for pasta phases appearing in the inner crust, and then
judge the crust-core transition by comparing the energy densities between
pasta phases and homogeneous matter.
Since the same nuclear model is employed for the description of the two
phases, the crust-core transition is determined in a consistent manner
and the resulting unified EOS is quite smooth.
In the present work, the Thomas--Fermi approximation is used only for
the inner crust but not for the outer crust. This is because the shell
effect is not considered within the Thomas--Fermi approximation.
In fact, when the Baym-Pethick-Sutherland (BPS) EOS~\cite{BPS71} for the outer
crust is replaced by the one obtained from the Thomas--Fermi calculation,
no significant difference is observed in the star radius.
Therefore, we prefer to use the BPS EOS for the outer crust below the
neutron drip density in the present calculations.

The recent GW170817 event triggered extensive studies for constraining
the EOS from measurements of the tidal deformability in a binary neutron-star
system~\cite{Zhang18,Tews18,Zhu18,Anna18,Lim18,Radi18,De18,Fatt18,Mali18,Dexh19,LiBA19}.
The analysis of GW170817 data provides valuable constraints on the tidal deformabilities
of the binary neutron-star merger~\cite{Abbo17,Abbo18}.
The correlation between the symmetry energy and the tidal deformability
was recently investigated within various frameworks~\cite{Fatt18,Mali18,Dexh19,LiBA19}.
In the present work, we use a set of unified EOSs to compute the tidal
deformability of neutron stars and study its dependence on the symmetry energy
slope $L$. Furthermore, we match different crust and core segments in order to
examine their influence on the resulting tidal deformability.

This article is organized as follows. In Sec.~\ref{sec:2},
we briefly describe the RMF model and the self-consistent Thomas--Fermi approximation
used for constructing the EOS.
In Sec.~\ref{sec:3}, we show the effects of the symmetry energy on neutron-star
properties using the unified EOS. Furthermore, the influences from the crust and
core segments are examined separately using two sets of non-unified EOSs.
Section~\ref{sec:4} is devoted to the conclusions.

\section{ Formalism}
\label{sec:2}

\begin{figure}[htbp]
 \begin{center}
  \includegraphics[clip,width=8.5 cm]{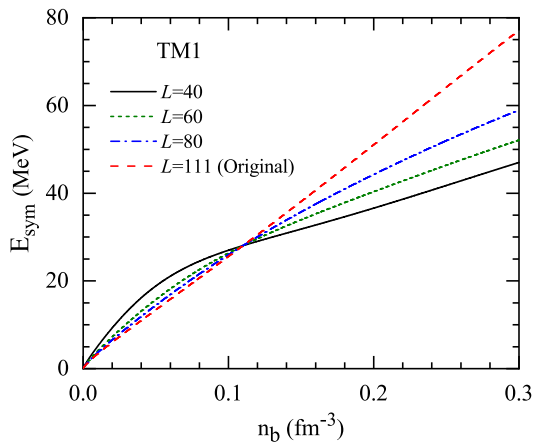}
  \caption{(Color online) Symmetry energy $E_{\rm{sym}}$ as a function of
  the baryon density $n_{b}$ for the generated TM1 models with different slope
  parameter $L$. The symmetry energy is fixed at a density of $0.11\, \rm{fm}^{-3}$.}
  \label{fig:1ESYM}
\end{center}
\end{figure}

We construct the EOS of neutron-star matter employing the RMF model for nuclear
interactions. In the RMF approach~\cite{Sero86,Meng06,TM1}, nucleons interact through
the exchange of various mesons, including the isoscalar-scalar meson $\sigma$,
the isoscalar-vector meson $\omega$, and the isovector-vector meson $\rho$.
For a system consisting of neutrons, protons, electrons, and muons,
the Lagrangian density reads
\begin{eqnarray}
\label{eq:LRMF}
\mathcal{L}_{\rm{RMF}} & = & \sum_{b=p,n}\bar{\psi}_b
\left\{i\gamma_{\mu}\partial^{\mu}-\left(M+g_{\sigma}\sigma\right) \right. \notag \\
&& \left. -\gamma_{\mu} \left[g_{\omega}\omega^{\mu} +\frac{g_{\rho}}{2}\tau_a\rho^{a\mu}
+\frac{e}{2}\left(1+\tau_3\right)A^{\mu}\right]\right\}\psi_b  \notag \\
&& +\frac{1}{2}\partial_{\mu}\sigma\partial^{\mu}\sigma -\frac{1}{2}%
m^2_{\sigma}\sigma^2-\frac{1}{3}g_{2}\sigma^{3} -\frac{1}{4}g_{3}\sigma^{4} \notag \\
&& -\frac{1}{4}W_{\mu\nu}W^{\mu\nu} +\frac{1}{2}m^2_{\omega}\omega_{\mu}%
\omega^{\mu} +\frac{1}{4}c_{3}\left(\omega_{\mu}\omega^{\mu}\right)^2
\notag \\
&& -\frac{1}{4}R^a_{\mu\nu}R^{a\mu\nu} +\frac{1}{2}m^2_{\rho}\rho^a_{\mu}%
\rho^{a\mu} -\frac{1}{4} F_{\mu\nu}F^{\mu\nu} \notag \\
&& +\Lambda_{\rm{v}} \left(g_{\omega}^2 \omega_{\mu}\omega^{\mu}\right)
\left(g_{\rho}^2\rho^a_{\mu}\rho^{a\mu}\right) \notag \\
&& +\sum_{l=e,\mu}\bar{\psi}_{l}\left(i\gamma_{\mu}\partial^{\mu} -m_{l} +e \gamma_{\mu}
A^{\mu} \right)\psi_{l}  ,
\end{eqnarray}
where $W^{\mu\nu}$, $R^{a\mu\nu}$, and $F^{\mu\nu}$ are the antisymmetric field
tensors corresponding to $\omega^{\mu}$, $\rho^{a\mu}$, and $A^{\mu}$, respectively.
In a static system within the mean-field approximation, the nonvanishing meson mean fields
are $\sigma =\left\langle \sigma \right\rangle$, $\omega =\left\langle\omega^{0}\right\rangle$,
$\rho =\left\langle \rho^{30} \right\rangle$, and $A =\left\langle A^{0}\right\rangle$.
The chemical potentials of nucleons are given by
\begin{eqnarray}
\mu _{p} &=&{\sqrt{\left( k_{F}^{p}\right)^{2}+{M^{\ast }}^{2}}}+g_{\omega}\omega
           +\frac{g_{\rho }}{2}\rho +e A,
\label{eq:mup} \\
\mu _{n} &=&{\sqrt{\left( k_{F}^{n}\right)^{2}+{M^{\ast }}^{2}}}+g_{\omega}\omega
           -\frac{g_{\rho }}{2}\rho ,
\label{eq:mun}
\end{eqnarray}%
where $M^{\ast}=M+g_{\sigma}\sigma$ is the effective nucleon mass
and $k_{F}^{i}$ is the Fermi momentum of species $i$, which is related
to the number density by $n_i=\left(k_{F}^{i}\right)^3/3\pi^2$.
It is noteworthy that the $\omega$-$\rho$ coupling term plays an important role in
determining the density dependence of the symmetry energy~\cite{Mene11,Prov13,Bao14b,IUFSU}.
The symmetry energy of nuclear matter is expressed as
\begin{eqnarray}
E_{\rm{sym}} &=& \frac{1}{2}{\left[\frac{\partial^2\left(\varepsilon/n_b\right)}
{\partial\alpha^2}\right]}_{\alpha=0} \notag \\
&=&
\frac{k^2_F}{6\sqrt{k^2_F+{M^\ast}^2}}+\frac{g^2_\rho n_b}
{8\left(m^2_{\rho}+2\Lambda_{\rm{v}}{g^2_\rho}{g^2_\omega}{\omega}^2\right)},
\end{eqnarray}
with $\alpha=\left(n_n-n_p\right)/n_b$ being the asymmetry parameter.
The slope of the symmetry energy is given by
\begin{equation}
L=3n_0\left[\frac{\partial E_{\rm{sym}}\left( n_b \right)}{\partial{n_b}}\right]_{n_b=n_0}.
\end{equation}
We use a set of generated models based on the TM1 parametrization~\cite{Bao14b},
where the coupling constants, $g_{\rho}$ and ${\Lambda}_{\rm{v}}$, are simultaneously
adjusted so as to achieve a given symmetry energy slope $L$ at saturation density $n_0$
while keeping the symmetry energy $E_{\rm{sym}}$ fixed at a density of $0.11\, \rm{fm}^{-3}$.
It was shown in Ref.~\cite{Bao14b} that all models in the set could provide the same
isoscalar properties and similar binding energies of finite nuclei as the
original TM1 model, but have different symmetry energy slope $L$.
To make the paper self-contained, we list in Table~\ref{tab:1}
the model parameters and saturation properties, while the calculated properties of
$^{208}$Pb are shown in the last three lines.
It is found that the models with different $L$ predict very similar binding energy
per nucleon and charge radius for $^{208}$Pb, whereas the neutron-skin thickness
$\triangle r_{\rm{np}}$ ($^{208}$Pb) obviously increases with increasing $L$.
We show in Fig.~\ref{fig:1ESYM} the symmetry energy $E_{\rm{sym}}$
as a function of the baryon density $n_b$ for all models listed in Table~\ref{tab:1}.
It is seen that the set of models has the same $E_{\text{sym}}$
at a density of $0.11\, \rm{fm}^{-3}$, but different values of
$E_{\text{sym}}$ at lower and higher densities due to different slope $L$.
The behavior of $E_{\text{sym}}$ plays a crucial role in determining
several properties of neutron stars.
\begin{table*}[htb]
\caption{Parameter sets used in this work and corresponding saturation properties.
The quantities $E_0$, $K$, $E_{\text{sym}}$, and $L$ are, respectively,
the energy per nucleon, incompressibility coefficient, symmetry
energy, and symmetry energy slope at saturation density $n_0$.
The last three lines show the neutron-skin thickness $\triangle r_{\text{np}}$,
charge radius $r_{\text{c}}$, and binding energy per nucleon $E/A$ of $^{208}$Pb.
The models are generated from the original TM1 model ($L$=111) by tuning
$g_\rho$ and $\Lambda_{\text{v}}$
to achieve a given slope $L$ at $n_0$ and a fixed symmetry energy
$E_{\text{sym}}=28.05$ MeV at a density of $0.11\, \rm{fm}^{-3}$.
Differences among these models are shown in bold.
Nucleon and meson masses are given in Refs.~\cite{TM1,Bao14b}. }
\label{tab:1}
\begin{center}
\begin{tabular}{lcccc}
\hline\hline
Model               & TM1($L$=40)  & TM1($L$=60)   & TM1($L$=80)   & TM1($L$=111) \\
\hline
$g_\sigma$          & 10.0289 & 10.0289 & 10.0289 & 10.0289  \\
$g_\omega$          & 12.6139 & 12.6139 & 12.6139 & 12.6139  \\
$g_{2}$ (fm$^{-1}$) & -7.2325 & -7.2325 & -7.2325 & -7.2325  \\
$g_{3}$             &  0.6183 &  0.6183 &  0.6183 &  0.6183  \\
$c_{3}$             & 71.3075 & 71.3075 & 71.3075 & 71.3075  \\
$g_\rho$            &  {\bf 13.9714} & {\bf 11.2610} & {\bf 10.1484} & {\bf 9.2644}  \\
$\Lambda_{\rm{v}}$  &  {\bf 0.0429}  & {\bf 0.0248}  & {\bf 0.0128}  & {\bf 0.0000}  \\
\hline
$n_0$ (fm$^{-3}$)   & 0.145   & 0.145 & 0.145    & 0.145   \\
$E_0$ (MeV)         & -16.3   & -16.3 & -16.3    & -16.3   \\
$K$ (MeV)           & 281     & 281   & 281      & 281     \\
$E_{\rm{sym}}$ (MeV)& {\bf 31.38} & {\bf 33.29}  & {\bf 34.86}  & {\bf 36.89}    \\
$L$ (MeV)           & {\bf 40}    & {\bf 60}     & {\bf 80}     & {\bf 111}     \\
\hline
$\triangle r_{\rm{np}}$ ($^{208}$Pb) (fm)  & {\bf 0.16}  & {\bf 0.21} & {\bf 0.24} & {\bf 0.27}  \\
$r_{\rm{c}}$ ($^{208}$Pb) (fm)     & {\bf 5.56} & {\bf 5.55} & {\bf 5.54}  & {\bf 5.54}  \\
$E/A$ ($^{208}$Pb) (MeV)           & 7.88  & 7.88 & 7.88   & 7.88    \\
\hline\hline
\end{tabular}%
\end{center}
\end{table*}

The $npe\mu$ matter in the neutron-star core has a uniform density
distribution under the conditions of $\beta$ equilibrium and charge neutrality.
The dense core EOS can be achieved
by solving a set of coupled equations in the RMF model. As the density decreases to about $n_0/2$,
the crust-core transition occurs, where nucleons cluster into pasta phases or spherical nuclei.
This is because the uniform matter is energetically unstable against cluster formation
at low densities. For nonuniform matter in the inner crust, we perform a self-consistent
Thomas--Fermi calculation as described in our previous work~\cite{Bao15}.
The Wigner--Seitz cell approximation is adopted to simplify the calculation of pasta phases.
The stable cell shape, which is determined by minimizing the energy density at a given density
$n_b$, may change from droplet to rod, slab, tube, and bubble as the density increases.
For simplicity, we assume the electron density is uniform throughout the
Wigner--Seitz cell. In the Thomas--Fermi approximation, the total energy per cell is calculated from
\begin{equation}
E_{\rm{cell}}=\int_{\rm{cell}}{\varepsilon}_{\rm{rmf}}({\bf r})d {\bf r}
  +{\varepsilon}_e V_{\rm{cell}},
\label{eq:TFe}
\end{equation}%
where ${\varepsilon}_{\rm{rmf}}({\bf r})$ is the local energy density at
position ${\bf r}$ given in the RMF model and ${\varepsilon}_e$ is the
kinetic energy density of electrons.
We consider different pasta configurations including the droplet, rod, slab, tube,
and bubble. The volume of the Wigner--Seitz cell for different configurations
is expressed as
\begin{equation}
V_{\rm{cell}}=\left\{
\begin{array}{ll}
\frac{4}{3}{\pi}r_{\text{ws}}^{3} &\hspace{0.5cm} \textrm{(droplet and bubble)},  \\
l{\pi}r_{\text{ws}}^{2}           &\hspace{0.5cm} \textrm{(rod and tube)},        \\
2r_{\text{ws}}{l^2}               &\hspace{0.5cm} \textrm{(slab)},
\end{array}
\right.
\label{eq:vcell}
\end{equation}
where $r_{\text{ws}}$ is the radius of a spherical cell for the droplet
and bubble configurations, while the rod and tube have cylindrical shapes
with radius $r_{\text{ws}}$ and length $l$ and the slab has width $l$
and thickness $2r_{\text{ws}}$. Actually, the energy density of the system
would not be affected by the choices of the length for a cylindrical shape
and the width for a slab.
At a given average baryon density $n_{b}$, we minimize the total
energy density $\varepsilon=E_{\rm{cell}}/V_{\rm{cell}}$ with respect to
the cell size $r_{\rm{ws}}$ for each pasta configuration, and then determine
the energetically favored state with the lowest energy density.
The pressure is calculated from the thermodynamic relation
\begin{eqnarray}
P=\sum_{i=b,l} \mu_{i} n_{i} - \varepsilon.
\label{eq:Pre}
\end{eqnarray}
The crust-core transition occurs at the density where the energy density
of the homogeneous phase becomes lower than that of the pasta phase.
It is well known that the symmetry energy slope $L$ plays an important role
in determining the pasta phase structure and the crust-core
transition~\cite{Gril12,Bao15,Oyam07}.
In Table~\ref{tab:2}, we present the onset densities of various nonspherical
nuclei and homogeneous matter for the generated TM1 models with different $L$.
It is seen that as $L$ increases, the crust-core transition density
(i.e., the onset density of homogeneous matter) significantly decreases
and some pasta phases disappear.
The model with $L$=40 MeV predicts the transition from droplet to rod occurs
at $n_b \approx 0.049\, \rm{fm}^{-3}$, then the pasta phases of slab, tube, and bubble
appear one by one, and finally transition to homogeneous matter occurs
at $n_b \approx 0.099\, \rm{fm}^{-3}$.
For the original TM1 model with $L$=111 MeV, only the droplet configuration
appears in the inner crust, and the transition from droplet to homogeneous matter
occurs at $n_b \approx 0.062\, \rm{fm}^{-3}$.
We note that the inner crust are calculated in the Thomas--Fermi approximation
for the density region between the neutron drip and the crust-core transition.
For the outer crust, we use the well-known BPS EOS, which is matched
to the inner-crust EOS at the neutron drip density.
\begin{table}[htb]
\caption{Onset densities given in the unit of fm$^{-3}$ for various nonspherical
nuclei (rod, slab, tube, and bubble) and homogeneous matter (HM)
obtained in the generated TM1 models with different $L$.}
\label{tab:2}
\begin{center}
\begin{tabular}{lccccc}
\hline\hline
Model               & Rod     & Slab    & Tube    & Bubble  & HM      \\
\hline
TM1($L$=40)         & 0.049   & 0.064   & 0.082   & 0.089   & 0.099   \\
TM1($L$=60)         & 0.066   & 0.076   & 0.081   &         & 0.083   \\
TM1($L$=80)         &         &         &         &         & 0.072   \\
TM1($L$=111)        &         &         &         &         & 0.062   \\
\hline\hline
\end{tabular}%
\end{center}
\end{table}

We apply the EOS constructed above to calculate the mass and radius
of a neutron star by solving the Tolman-Oppenheimer-Volkoff (TOV) equation in units of $G=c=1$,
\beq
\label{eq:tov}
\frac{dP(r)}{dr}&=&-\frac{M(r)\varepsilon(r)}{r^{2}}\left[1+\frac{P(r)}{\varepsilon(r)}\right] \notag \\
&& \times \left[1+\frac{4\pi r^{3}P(r)}{M(r)}\right]
\left[1-\frac{2M(r)}{r}\right]^{-1}, \\
\frac{dM(r)}{dr}&=&4\pi r^{2}\varepsilon(r),
\eeq
where $P(r)$ and $\varepsilon(r)$ are the pressure and energy density at the radial
coordinate $r$, respectively. $M(r)$ is the gravitational mass enclosed within the radius $r$.
The dimensionless tidal deformability of a neutron star is expressed as~\cite{Hind08,Post10}
\beq
\label{eq:td}
\Lambda=\frac{2}{3}k_2C^{-5},
\eeq
where $C=M/R$ is the compactness parameter of the star with mass $M$ and radius $R$.
The tidal Love number $k_2$ is calculated from
\beq
k_2&=&\frac{8C^5}{5}(1-2C)^2\left[2-y_R+2C(y_R-1)\right]\nn
&& \times\left\{2C\left[6-3y_R+3C(5y_R-8)\right]\right.\nn
&& +4C^3\left[13-11y_R+C(3y_R-2)+2C^2(1+y_R)\right]\nn
&& +3(1-2C)^2\left[2-y_R+2C(y_R-1)\right]\nn
&& \left.\times\ln(1-2C)\right\}^{-1},
\eeq
where $y_R=y(R)$ is obtained by solving the following differential equation:
\beq
r\frac{dy(r)}{dr}+y(r)^2+y(r)F(r)+r^2Q(r)=0,
\eeq
with
\beq
F(r)&=&\left\{1-4\pi r^2 [\varepsilon(r)-P(r)] \right\}\left[1-\frac{2M(r)}{r}\right]^{-1},\\
Q(r)&=&4\pi\left[5\varepsilon(r)+9P(r)+\frac{\varepsilon(r)+P(r)}
       {\partial P(r)/\partial\varepsilon(r)}-\frac{6}{4\pi r^2}\right] \nn
&& \times  \left[1-\frac{2M(r)}{r}\right]^{-1}
-\frac{4M(r)^2}{r^4}\left[1+\frac{4\pi r^3P(r)}{M(r)}\right]^2 \nn
&& \times \left[1-\frac{2M(r)}{r}\right]^{-2}.
\eeq
In a binary neutron-star system, the tidal effect is given by
the combined dimensionless tidal deformability
\beq
\label{eq:btd}
\tilde{\Lambda}=\frac{16}{13}
\frac{(12q+1)\Lambda_1+(12+q)q^4\Lambda_2}{(1+q)^{5}},
\eeq
where $\Lambda_1$ and $\Lambda_2$ are the individual tidal deformabilities
of the two neutron stars with the mass ratio $q=M_2/M_1 \leq 1$.

\section{Results and discussion}
\label{sec:3}

We present numerical results for neutron-star properties using the EOSs
obtained with the set of RMF models. To examine the effects of the symmetry
energy, we apply the unified EOS to compute various properties of neutron stars.
In order to separately investigate the influence of crust and core segments,
nonunified EOSs are used by matching different crust and core EOSs.

\begin{figure}[htbp]
 \begin{center}
  \includegraphics[clip,width=8.5 cm]{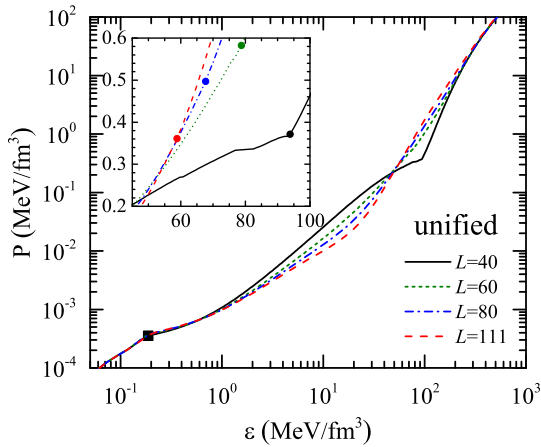}
  \caption{(Color online) Pressure $P$ as a function of the energy density $\varepsilon$
  obtained using the set of generated TM1 models with different $L$
  for the inner crust and core. The BPS EOS is adopted for the outer crust and
  the matching point is marked by the filled square.
  The crust-core transition is indicated by the filled circles in the inset.}
  \label{fig:2PE0}
 \end{center}
\end{figure}

\begin{figure}[htbp]
 \begin{center}
  \includegraphics[clip,width=8.5 cm]{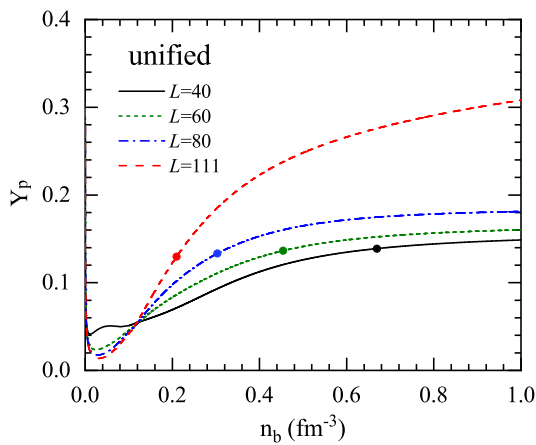}
  \caption{(Color online) Proton fraction $Y_p$ of the unified EOSs
  as a function of the baryon density $n_b$ for the set of generated
  TM1 models. The filled circles indicate the threshold for the dUrca process.}
  \label{fig:3YP0}
 \end{center}
\end{figure}

\begin{figure}[htbp]
 \begin{center}
  \includegraphics[clip,width=8.5 cm]{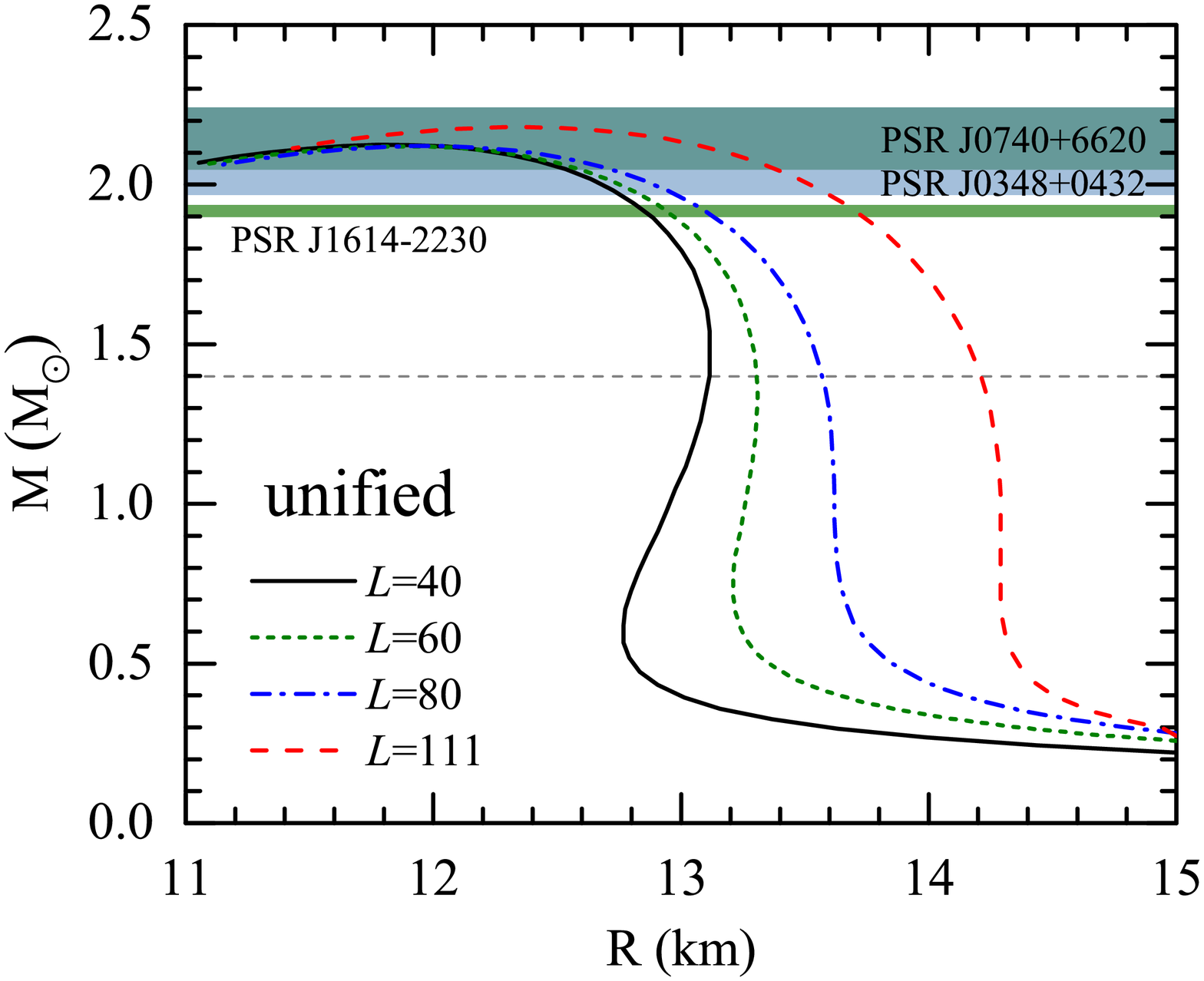}
  \caption{(Color online) Mass-radius relations of neutron stars obtained using the
  unified EOSs shown in Fig.~\ref{fig:2PE0}.
  The horizontal bars indicate the recent neutron-star mass measurements of
  PSR J1614--2230~\cite{Demo10,Fons16}, PSR J0348+0432~\cite{Anto13},
  and PSR J0740+6620~\cite{Crom19}.}
  \label{fig:4MR0}
 \end{center}
\end{figure}

\begin{figure*}[htbp]
  \includegraphics[clip,width=8.5 cm]{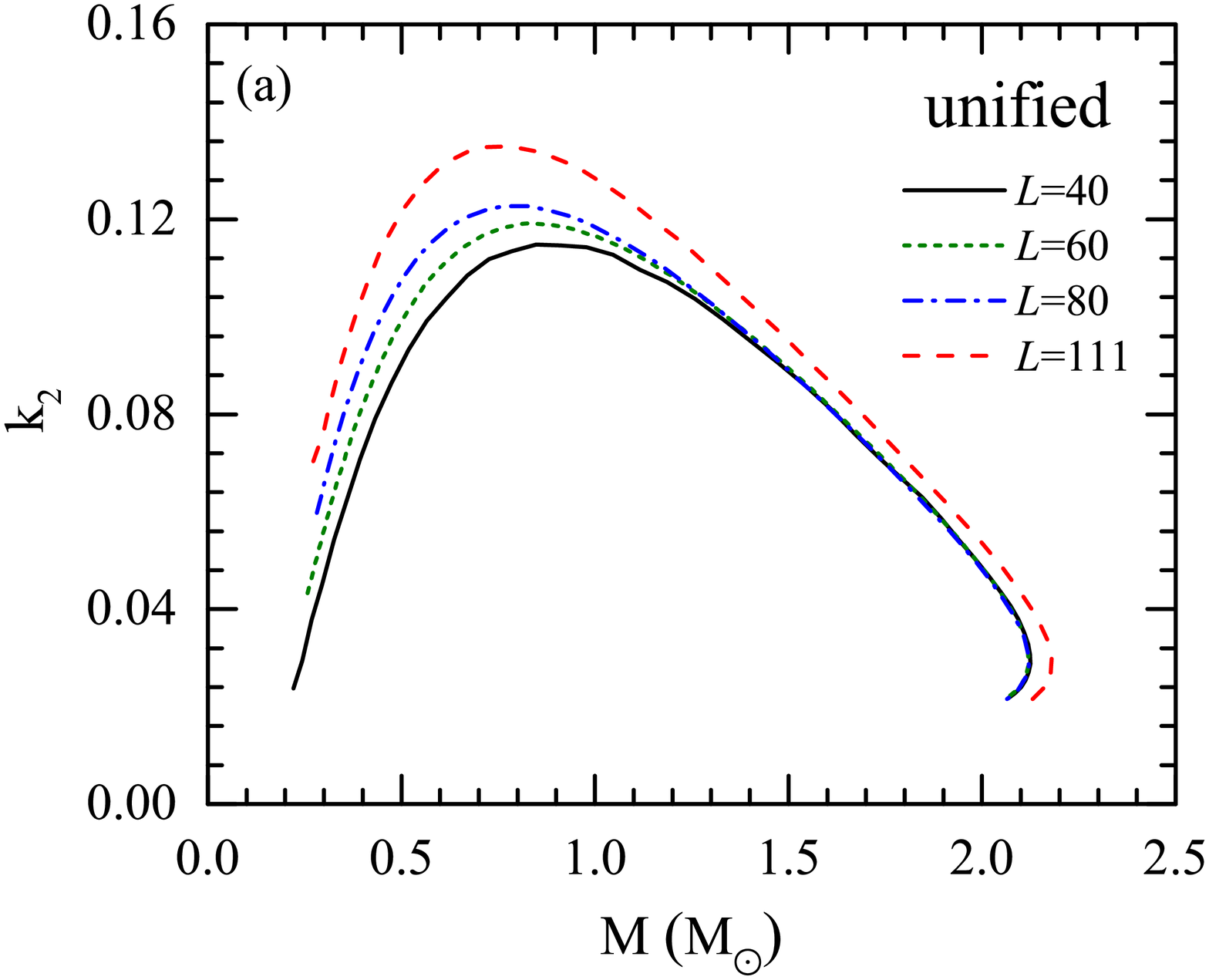}
  \includegraphics[clip,width=8.5 cm]{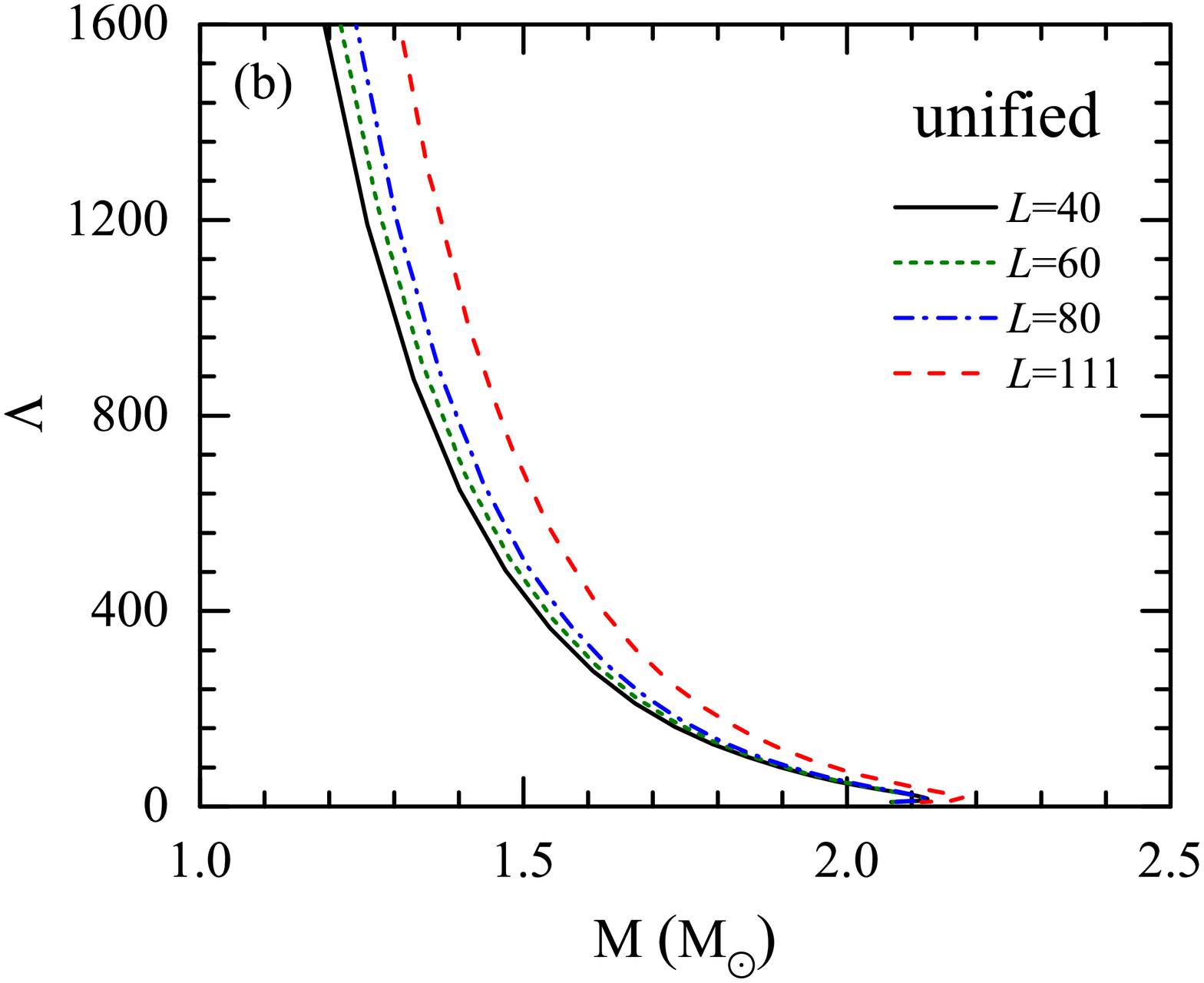}
  \caption{(Color online) Love number $k_{2}$ and tidal deformability $\Lambda$
  as a function of the neutron-star mass $M$ obtained using the
  unified EOSs shown in Fig.~\ref{fig:2PE0}.}
\label{fig:5KL0}
\end{figure*}

\begin{figure}[htbp]
 \begin{center}
  \includegraphics[clip,width=8.5 cm]{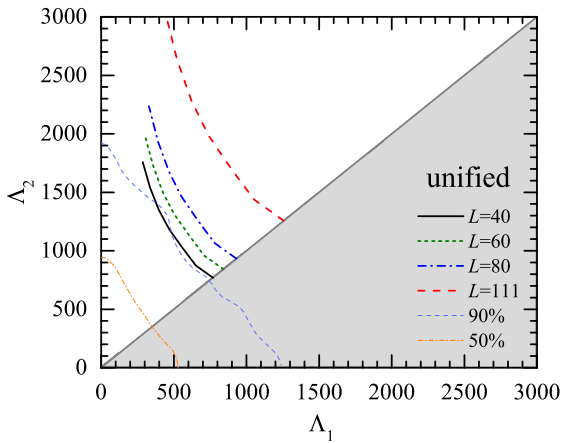}
  \caption{(Color online) Tidal deformabilities $\Lambda_1$ vs $\Lambda_2$
  of the two neutron stars in GW170817, using the unified EOSs
  with different slope parameters $L$.
  The $90\%$ and $50\%$ credible constraints from the latest analysis of GW170817~\cite{Abbo18}
  are shown by thin dashed and dash-dotted lines, respectively. }
  \label{fig:6LAM120}
 \end{center}
\end{figure}

\subsection{Neutron-star properties with unified EOSs}
\label{sec:3.1}
The unified EOS used in this work is obtained by performing a self-consistent
Thomas--Fermi calculation for the inner crust, which is smoothly connected to
the core EOS based on the same nuclear model. We use the BPS EOS for
the outer crust below the neutron drip density.
In Fig.~\ref{fig:2PE0}, we plot the pressure $P$ as a function of the energy
density $\varepsilon$ obtained using the set of generated TM1 models
with different slope parameters $L$.
The crust-core transition is indicated by the filled circles.
It is shown that the model with a small value of $L$ predicts a large crust-core
transition and relatively small pressures at high densities.
In the Thomas--Fermi approximation, the phase transition is determined by
minimizing the energy density. As a result, the energy density
is a smooth function of the baryon density, but the pressure as the first
derivative of the energy may exhibit a weak discontinuity of first-order phase
transition~\cite{Pais14}. In Fig.~\ref{fig:2PE0}, a clear kink in the TM1($L$=40) EOS
is observed at the crust-core transition, whereas it is invisible in other cases.
This is because the TM1($L$=40) EOS has relatively small pressure and large
crust-core transition density.

It is well known that the most efficient mechanism for neutron-star cooling
is the direct Urca (dUrca) process, i.e., the electron capture by a proton and the
beta decay of a neutron. The threshold for the dUrca process is mainly
determined by the proton fraction $Y_p$ in the cores of neutron stars, where
the proton fraction is large enough to allow for momentum conservation.
In simple $npe$ neutron-star matter, the dUrca process can occur for
$Y_p \geq 1/9$. When muons are included under the equilibrium condition
$\mu_e=\mu_\mu$, the critical $Y_p$ for the dUrca process is
in the range of $(11.1-14.8)\%$~\cite{Latt91}.
In fact, the proton fraction $Y_p$ of neutron-star matter is strongly
dependent on the symmetry energy.
In Fig.~\ref{fig:3YP0}, the proton fraction $Y_p$ of the unified EOSs
is plotted as a function of the baryon density $n_b$ for the set of generated
TM1 models, and the corresponding threshold for the dUrca process
is indicated by the filled circles.
These models show different behaviors of the symmetry energy due to different slope
parameters $L$. The model with $L$=40 MeV predicts a small $Y_p$ at high densities
and a large threshold density of $0.67\, \rm{fm}^{-3}$ for the dUrca process.
In contrast, the original TM1 model ($L$=111 MeV) gives a much higher $Y_p$ and
small threshold density of $0.21\, \rm{fm}^{-3}$.
It has been reported in Ref.~\cite{Negr18} that neutron-star cooling observations
are more compatible with an EOS having a smaller value of $L$.
Therefore, the TM1($L$=40) model is more favored by the cooling
observations than the TM1($L$=111) model.

We present, in Fig.~\ref{fig:4MR0}, the resulting mass-radius relation with the set
of unified EOSs. It is found that the maximum mass of neutron stars lies in the range
of $2.12-2.18 M_\odot$, which is compatible with the observational constraints of
PSR J1614--2230 ($M=1.928 \pm 0.017  M_\odot$)~\cite{Demo10,Fons16},
PSR J0348+0432 ($M=2.01  \pm 0.04   M_\odot$)~\cite{Anto13}, and
PSR J0740+6620 ($M=2.14^{+0.10}_{-0.09}  M_\odot$)~\cite{Crom19}.
It is shown that the maximum mass is not very sensitive to the slope parameter $L$,
but the radius obviously depends on the value of $L$. We find that the radius of
a canonical $1.4 M_\odot$ neutron star ($R_{1.4}$) is $\approx 14.21$ km using the TM1($L$=111) model,
while it reduces to $\approx 13.12$ km with the TM1($L$=40) model.
So far, the precise measurement of neutron-star radii is still a challenge
for astrophysical observations, and no stringent constraints on the radius $R_{1.4}$
can be derived~\cite{Latt14,Haen16}.
The recent analysis of GW170817 data provides a constraint on the radius of a
$1.4 M_\odot$ neutron star of $R_{1.4}<13.6$ km~\cite{Abbo17}.
Many studies based on different approaches for the GW170817 event suggested a consistent
upper limit for the radius of a $1.4 M_\odot$ neutron star
as $R_{1.4}<13.8$ km~\cite{Tews18,Zhu18,De18,Fatt18,Mali18}.
Our resulting $R_{1.4}$ with a smaller $L$ is compatible with this constraint.
It is noteworthy that the calculations of neutron-star radii are model dependent,
as can be found in Refs.~\cite{Latt16,Oert17,Fort16,Gand12}.
In Ref.~\cite{Gand12}, quantum Monte Carlo calculations predict $R_{1.4}<12$ km
for $L\leq 45$ MeV, which are much smaller than our results.
Therefore, the slope parameter $L$ cannot be precisely constrained by observations
of neutron-star radii due to the model dependency.
On the other hand, the positive correlation between $L$ and $R_{1.4}$
is consistent among different models.

It is interesting to examine the correlation between the tidal deformability of
neutron stars and the density dependence of nuclear symmetry energy.
The tidal deformability is determined by the EOS through both the tidal Love
number $k_2$ and the compactness parameter $C=M/R$, as shown in Eq.~(\ref{eq:td}).
We plot in Fig.~\ref{fig:5KL0} the tidal Love number $k_2$ (left panel)
and the dimensionless tidal deformability $\Lambda$ (right panel) as a function of the neutron-star
mass $M$. One can see that $k_2$ increases with the neutron-star mass and reaches
its maximum value around 0.7--0.9 $M_\odot$, and then decreases rapidly in the large-mass region. We find that there are significant differences in $k_2$ for a fixed $M$
between the EOSs with different slope parameters $L$, especially for smaller
neutron-star masses.
The model with a small $L$ predicts a small value of $k_2$, and therefore
a small tidal deformability $\Lambda$ is achieved due to their relation in Eq.~(\ref{eq:td}).
It is shown that a clear $L$ dependence of the tidal deformability $\Lambda$
is observed, which comes from the $L$ dependence of both the tidal Love
number $k_2$ and the radius $R$. The value of $\Lambda$ is very large for
a small neutron-star mass due to its small compactness parameter.
As the star mass increases, the tidal deformability $\Lambda$ decreases rapidly.
For the canonical $1.4 M_\odot$ neutron star, we obtain $\Lambda=652$ using the
TM1($L$=40) model, while it increases to $\Lambda=1047$ for the TM1($L$=111) model.
The analysis of GW170817 data has placed a constraint on the tidal deformability of
a $1.4 M_\odot$ neutron star, i.e., $\Lambda_{1.4} < 800$~\cite{Abbo17}.
Hence, an EOS with a small symmetry energy slope like $L$=40 MeV is more favored than
one with a large slope like $L$=111 MeV.

In Fig.~\ref{fig:6LAM120}, we plot the tidal deformabilities $\Lambda_1$ vs $\Lambda_2$
of the two neutron stars in GW170817, using the unified EOSs
with different slope parameters. $\Lambda_1$ and $\Lambda_2$ are the
individual tidal deformabilities associated with the high-mass $M_1$ and low-mass $M_2$
components of the binary, respectively. The curves are obtained by varying
independently the high-mass component in the range $1.365 \leq M_1/M_\odot \leq 1.60$,
whereas the low-mass component is determined by keeping the chirp mass fixed at
the observed value of $\mathcal{M}=(M_1 M_2)^{3/5}(M_1+M_2)^{-1/5}=1.188 M_\odot$~\cite{Abbo17}.
The $90\%$ and $50\%$ credible constraints from the latest analysis of GW170817 by
LIGO and Virgo Collaborations~\cite{Abbo18} are shown by thin dashed and dash-dotted
lines, respectively. Compared to the $90\%$ confidence limit reported in the initial
analysis of GW170817~\cite{Abbo17}, the present $90\%$ credible constraint is
considerably reduced. We can see that the curve obtained by the TM1($L$=40) model
is compatible with the $90\%$ credible constraint, but other curves with larger $L$
are almost ruled out. The correlation between the tidal deformability and the slope parameter
suggests that large values of $L$ are not favored by GW170817.

\begin{figure}[htbp]
\begin{center}
  \includegraphics[clip,width=8.5 cm]{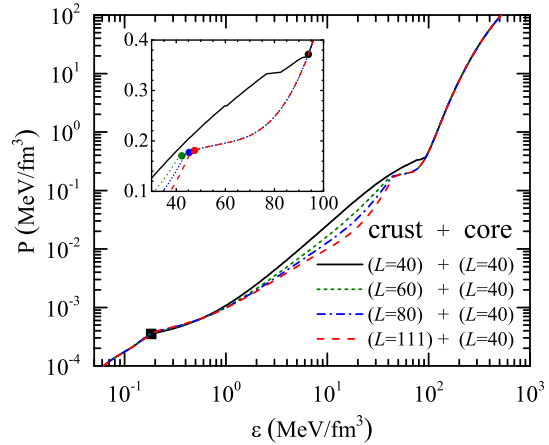}
  \caption{(Color online) Pressure $P$ as a function of the energy density $\varepsilon$
  obtained using the generated TM1 models with different $L$ for the inner crust and
  the TM1($L$=40) model for the core. The crust-core transition is indicated by the filled
  circles in the inset. The BPS EOS is adopted for the outer crust and
  the matching point is marked by the filled square. }
  \label{fig:7PE1}
\end{center}
\end{figure}

\begin{figure}[htbp]
 \begin{center}
  \includegraphics[clip,width=8.5 cm]{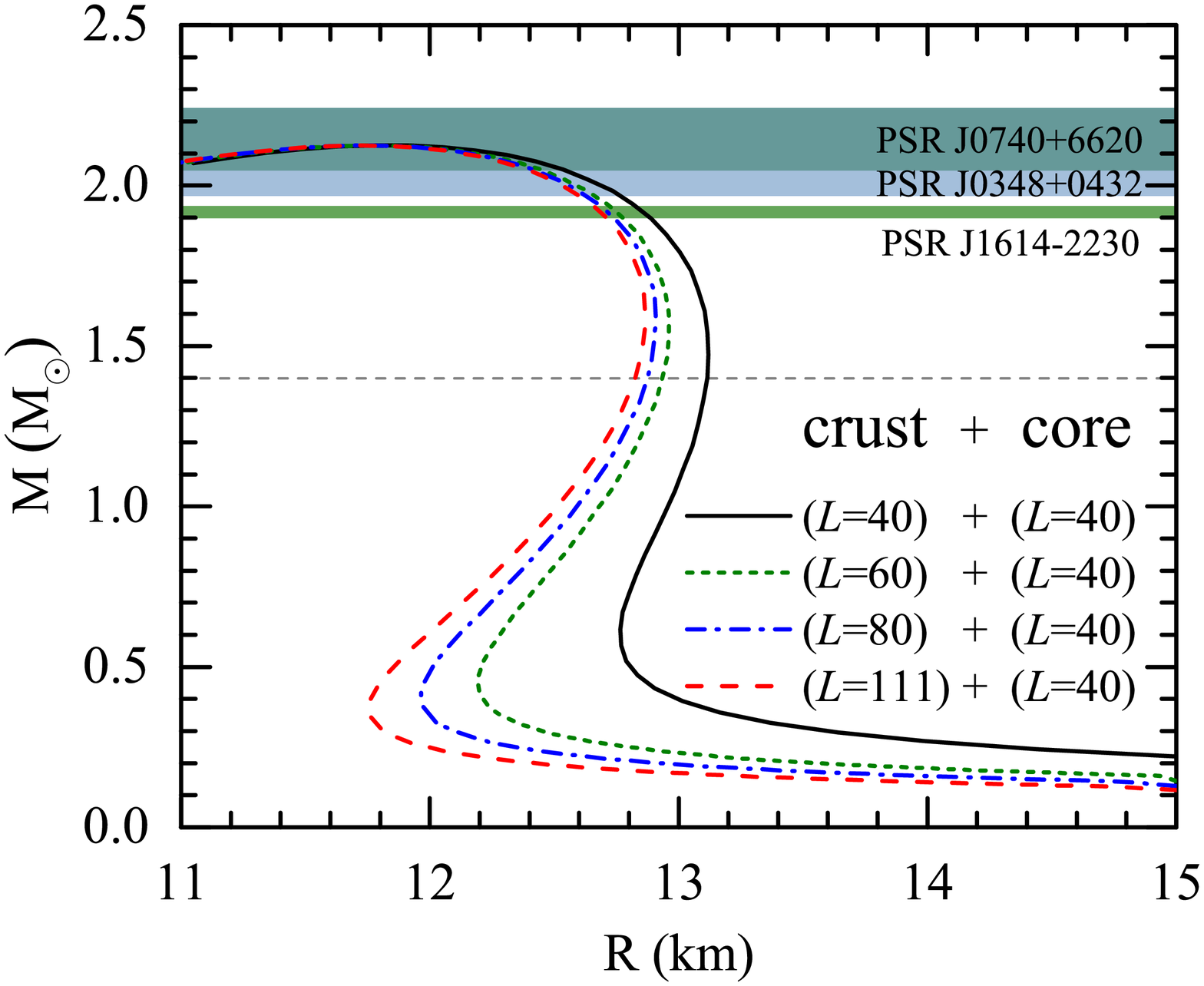}
  \caption{(Color online) Mass-radius relations of neutron stars obtained using the
  nonunified EOSs shown in Fig.~\ref{fig:7PE1}.
  The horizontal bars indicate the recent neutron-star mass measurements of
  PSR J1614--2230~\cite{Demo10,Fons16}, PSR J0348+0432~\cite{Anto13},
  and PSR J0740+6620~\cite{Crom19}. }
  \label{fig:8MR1}
 \end{center}
\end{figure}

\begin{figure*}[htbp]
  \includegraphics[clip,width=8.5 cm]{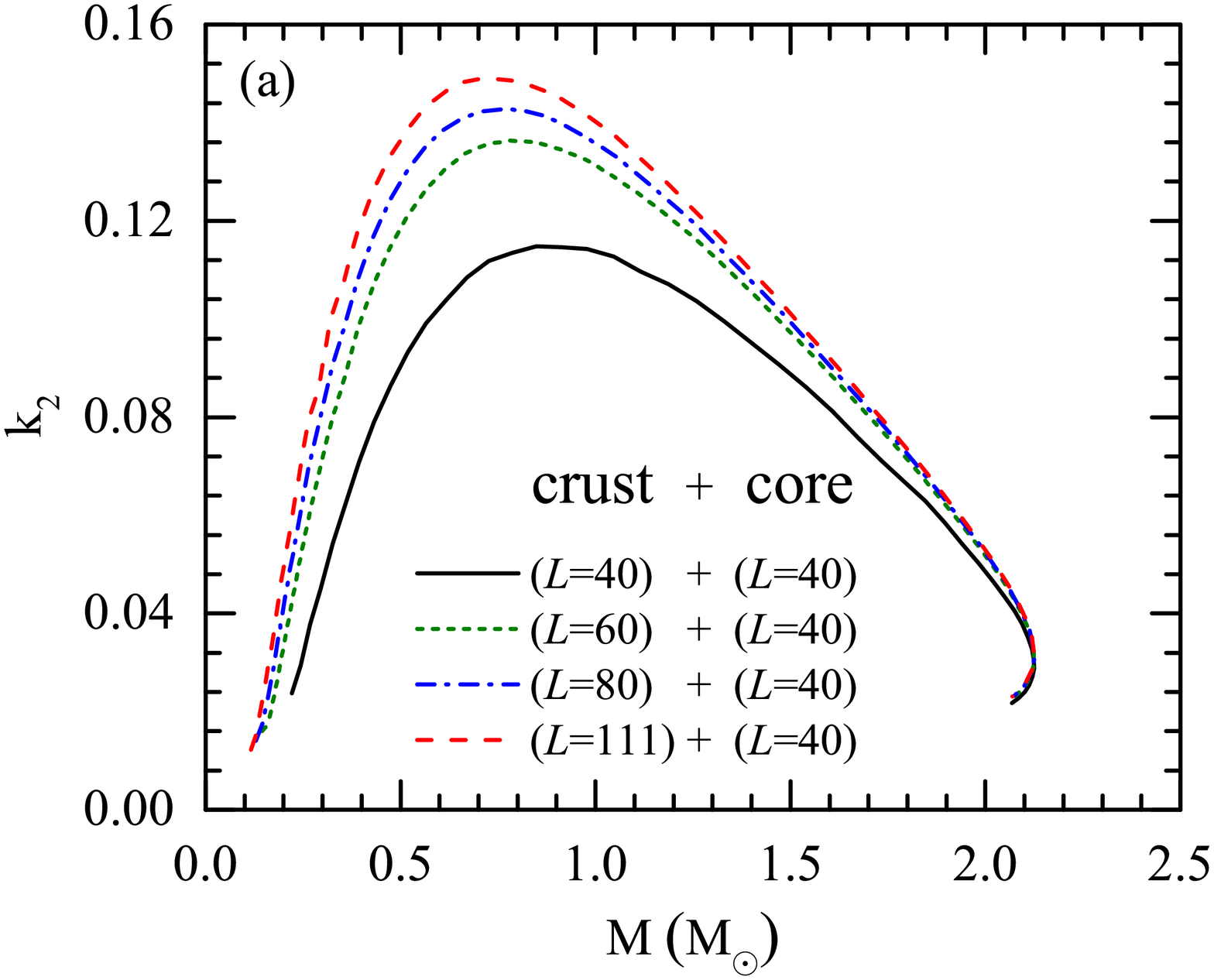}
  \includegraphics[clip,width=8.5 cm]{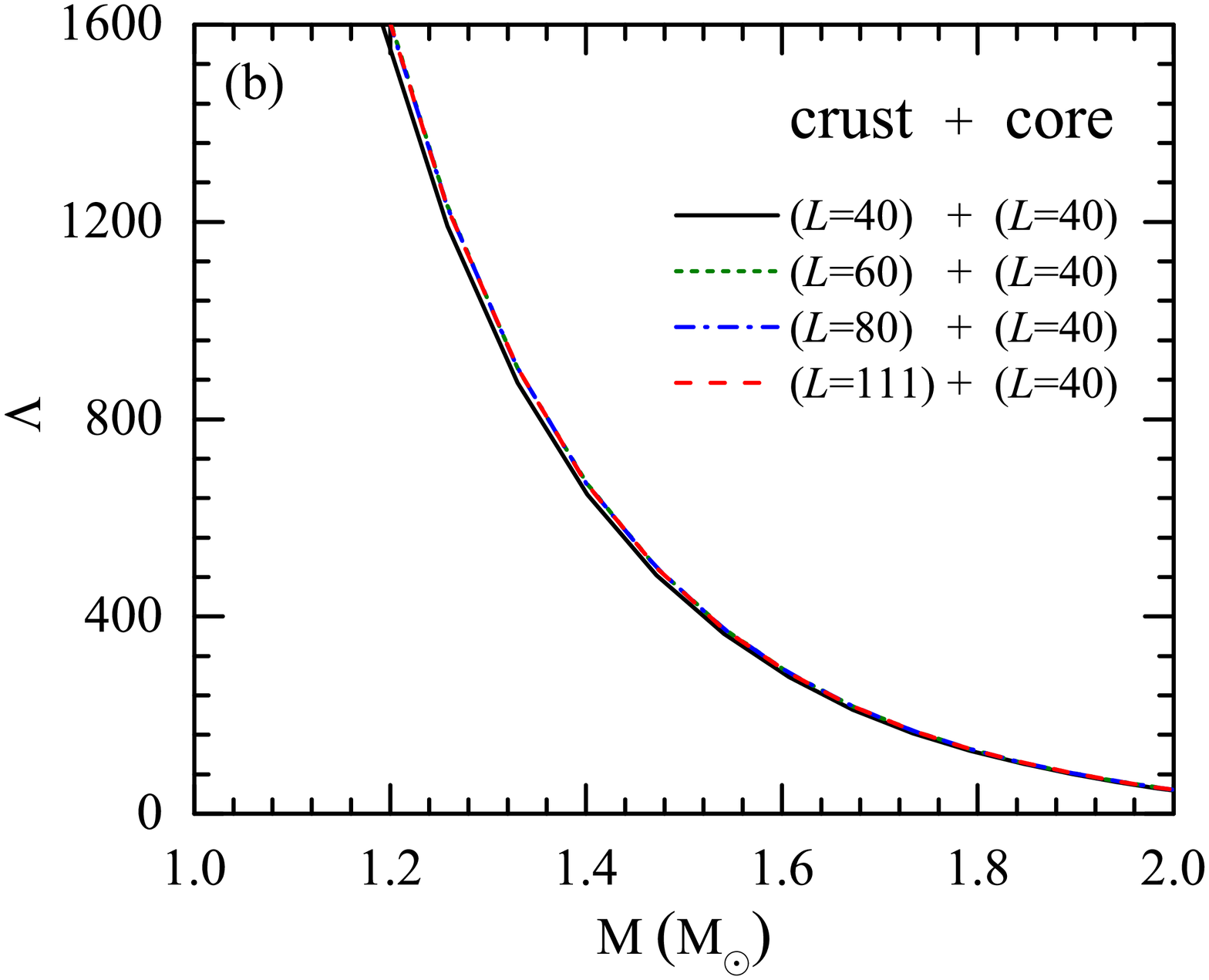}
  \caption{(Color online) Love number $k_{2}$ and tidal deformability $\Lambda$
  as a function of the neutron-star mass $M$ obtained using the
  nonunified EOSs shown in Fig.~\ref{fig:7PE1}. }
\label{fig:9KL1}
\end{figure*}

\begin{figure}[htbp]
 \begin{center}
  \includegraphics[clip,width=8.5 cm]{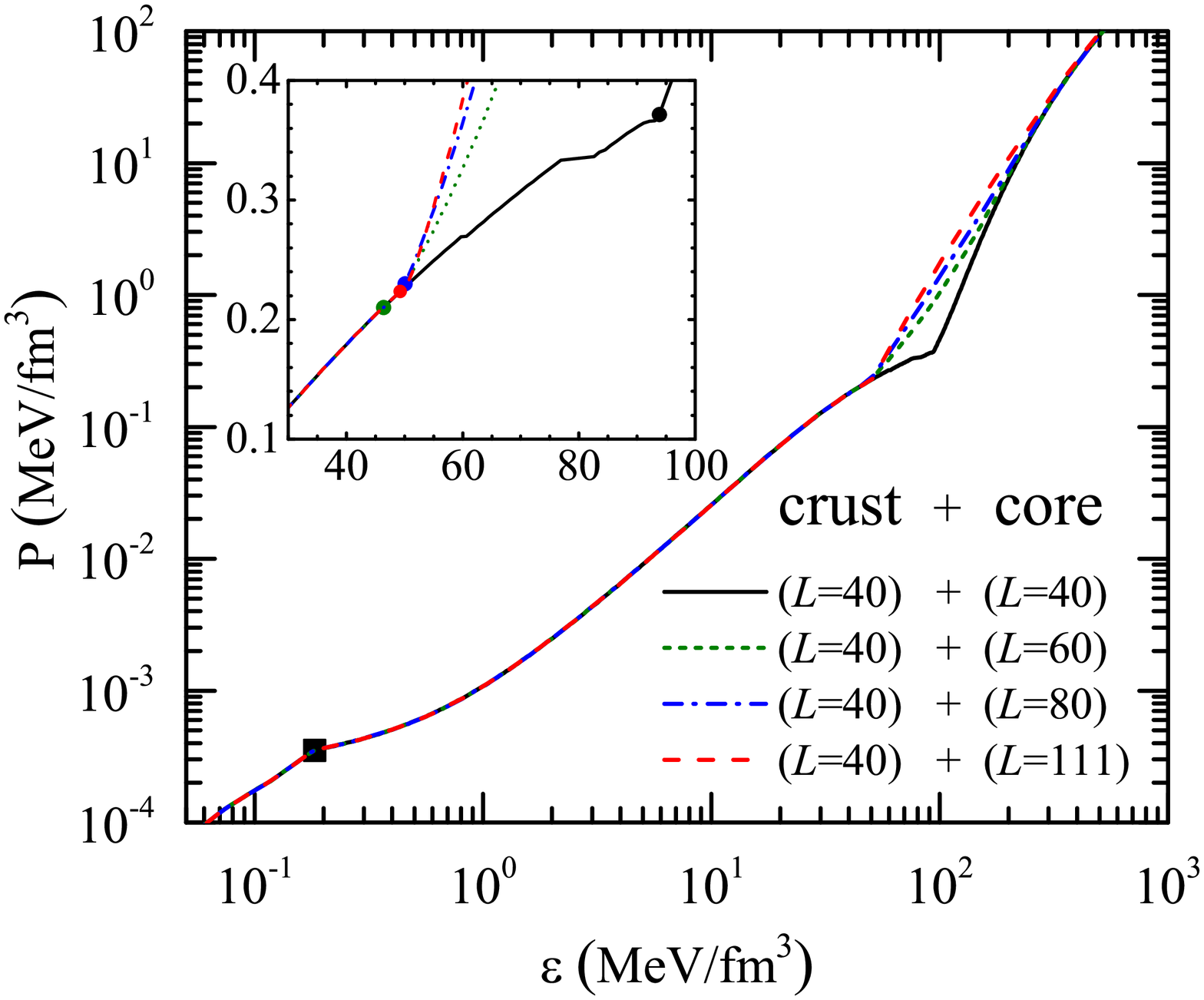}
  \caption{(Color online) Pressure $P$ as a function of the energy density $\varepsilon$
  obtained using the generated TM1 models with different $L$ for the core and
  TM1($L$=40) for the inner crust. The crust-core transition is indicated by the filled
  circles in the inset. The BPS EOS is adopted for the outer crust and
  the matching point is marked by the filled square. }
  \label{fig:10PE2}
 \end{center}
\end{figure}

\begin{figure}[htbp]
 \begin{center}
  \includegraphics[clip,width=8.5 cm]{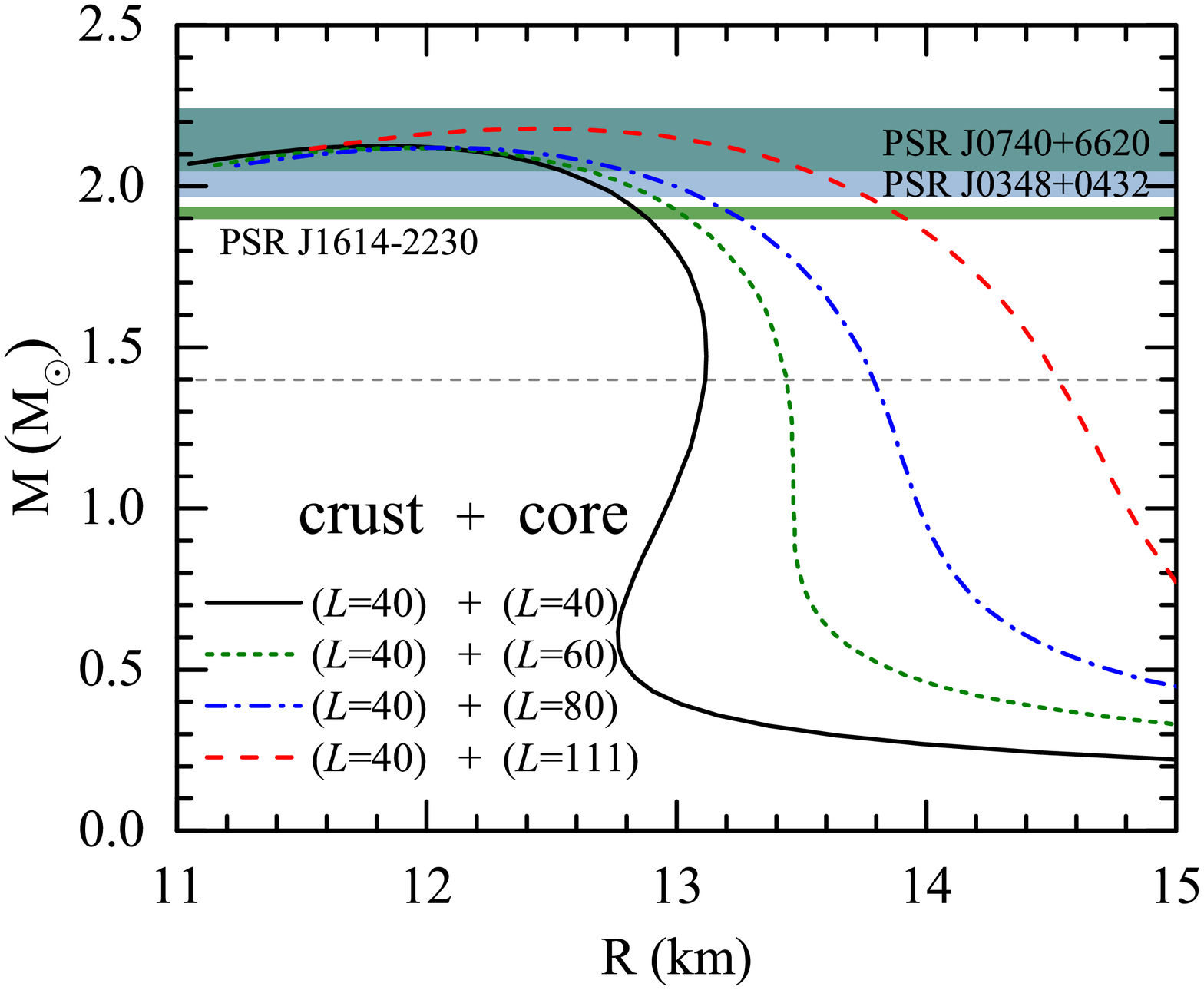}
  \caption{(Color online) Mass-radius relations of neutron stars obtained using the
  nonunified EOSs shown in Fig.~\ref{fig:10PE2}.
  The horizontal bars indicate the recent neutron-star mass measurements of
  PSR J1614--2230~\cite{Demo10,Fons16}, PSR J0348+0432~\cite{Anto13},
  and PSR J0740+6620~\cite{Crom19}. }
  \label{fig:11MR2}
 \end{center}
\end{figure}

\begin{figure*}[htbp]
  \includegraphics[clip,width=8.5 cm]{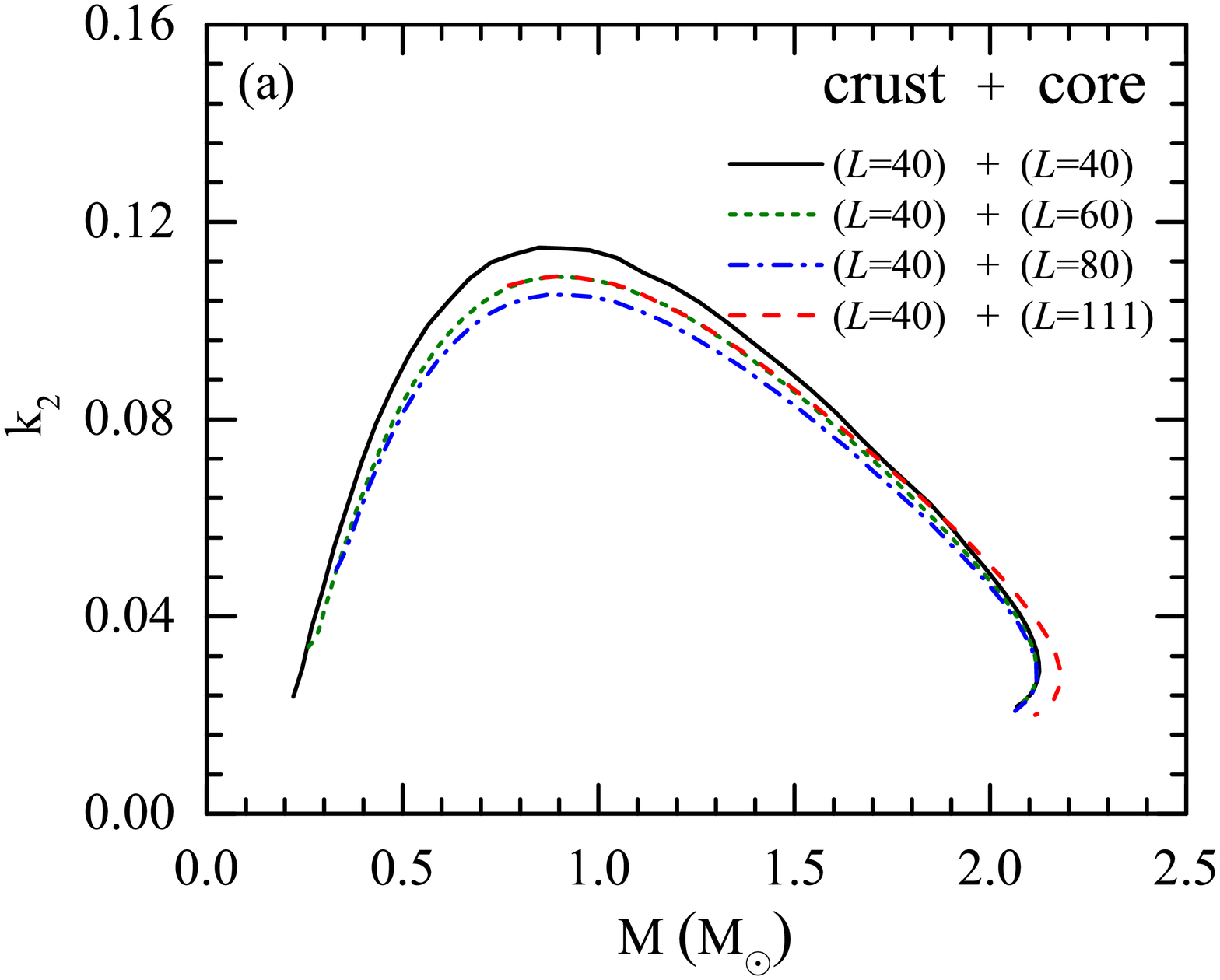}
  \includegraphics[clip,width=8.5 cm]{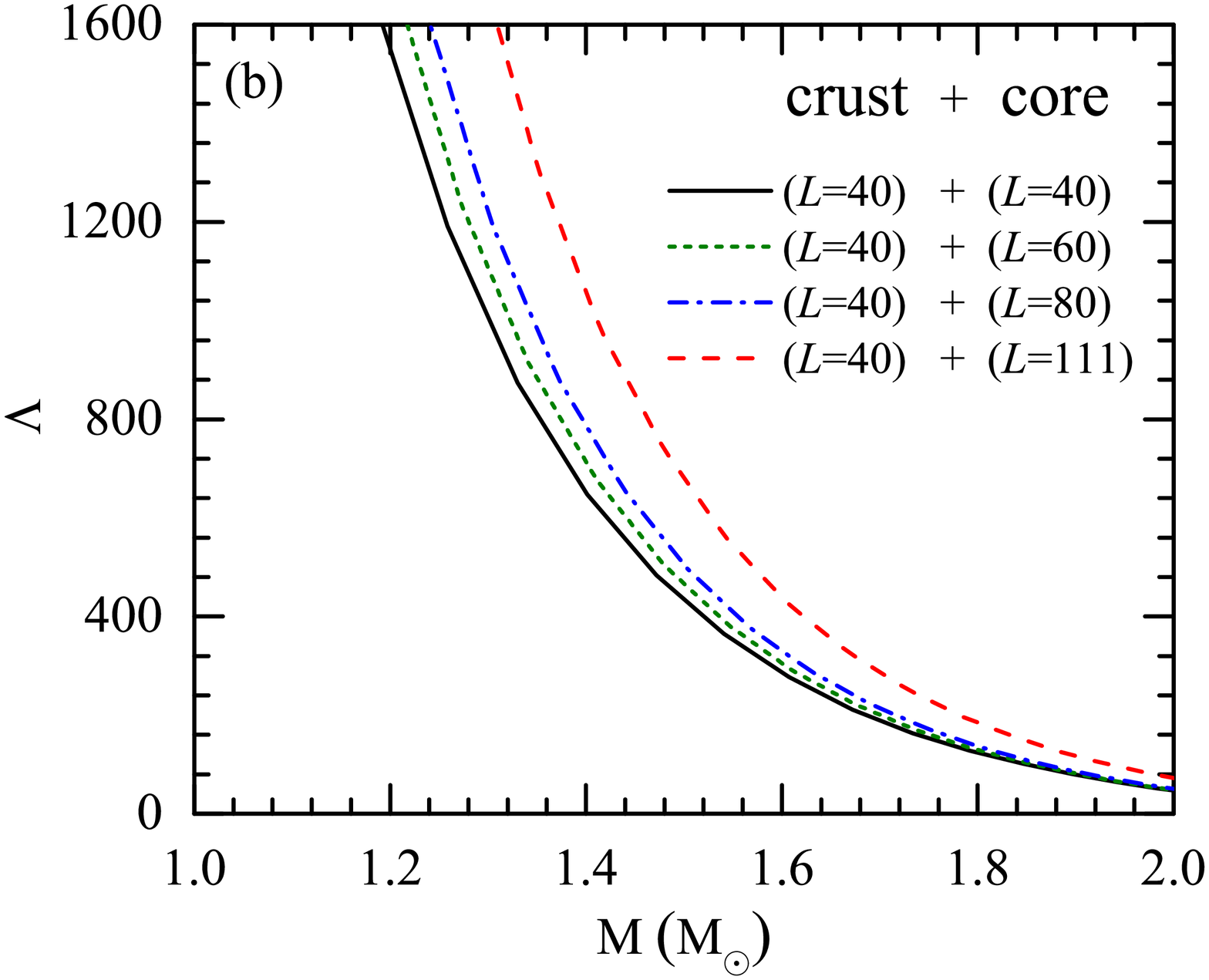}
  \caption{(Color online) Love number $k_{2}$ and tidal deformability $\Lambda$
  as a function of the neutron-star mass $M$ obtained using the
  nonunified EOSs shown in Fig.~\ref{fig:10PE2}. }
  \label{fig:12KL2}
\end{figure*}

\subsection{Effects of the crust EOS}
\label{sec:3.2}
We separately investigate the effects of crust and core EOSs on neutron-star properties.
To examine the effect of the crust, we construct a set of nonunified EOSs by matching
the same core EOS to different crust segments. The crust-core transition is determined by
the crossing point of the two segments, where the crust and core have equal pressure and
energy density. In Fig.~\ref{fig:7PE1}, we display the pressure $P$ as a function of
the energy density $\varepsilon$ for the set of nonunified EOSs, where the TM1($L$=40)
model is used for the core and the inner crust is described by the models with different
slope parameter $L$. It is shown that there are obvious differences in the inner crust region
among these EOSs, whereas no difference exists both in the outer crust BPS EOS
and in the core TM1($L$=40) EOS. The model with a large $L$ predicts a soft EOS of the inner crust,
which is opposite to the behavior at high densities (see Fig.~\ref{fig:2PE0}).
Therefore, the softest EOS considered here is the combination of the crust with $L$=111 MeV
and the core with $L$=40 MeV.
The $L$ dependence of the EOS can be understood from the density dependence of
the symmetry energy $E_{\rm{sym}}$ shown in Fig.~\ref{fig:1ESYM}.

It is interesting to examine quantitatively the effect of the inner crust on
neutron-star properties.
In Fig.~\ref{fig:8MR1}, we plot the mass-radius relation obtained using the set
of nonunified EOSs. It is noticed that almost no difference is found for massive
neutron stars when using different EOSs, which indicates the crust contribution
is unimportant for a large mass star. On the other hand, the difference in the radius
becomes more pronounced as the mass decreases.
For the canonical $1.4 M_\odot$ neutron star, the radius $R_{1.4}$ changes from
$\approx 13.12$ km using the unified TM1($L$=40) EOS to $\approx 12.82$ km when replacing
the crust EOS with TM1($L$=111). This means that the difference in the crust EOS
may lead to $\approx 0.3$ km difference in $R_{1.4}$.
Furthermore, it is found that a small $L$ of the crust corresponds to a large
neutron-star radius, which is opposite to the $L$ dependence shown in Fig.~\ref{fig:4MR0}.
This is because the model with a small $L$ results in a hard EOS at subnuclear densities
and a soft EOS at supernuclear densities. In the case of the unified EOS,
the neutron-star radius is determined dominantly by the core EOS, where
the crust EOS is less important. When the nonunified EOSs shown in Fig.~\ref{fig:7PE1}
are employed, the differences in the radii come only from the inner crust segments.
Therefore, the sensitivity of the radius to the crust EOS can be examined by using
this set of nonunified EOSs.

To study the influence of the crust EOS on the tidal deformability of
neutron stars, we show in Fig.~\ref{fig:9KL1} the tidal Love number $k_2$ (left panel)
and the dimensionless tidal deformability $\Lambda$ (right panel) as a function of
the neutron-star mass $M$, using the set of nonunified EOSs shown in Fig.~\ref{fig:7PE1}.
The behavior of $k_2$ in this case is very similar to that using unified EOSs,
as shown in the left panel of Fig.~\ref{fig:5KL0}.
The maximum values of $k_2$ obtained using the non-unified EOSs are somewhat higher
than corresponding results of unified EOSs. Although the same core EOS is adopted for
all non-unified EOSs considered, significant differences are found in $k_2$
due to the difference of the inner crust. This implies that the tidal Love number $k_2$
is rather sensitive to the crust EOS. However, the tidal deformability $\Lambda$ shown
in right panel of Fig.~\ref{fig:9KL1} is not so sensitive to the crust EOS.
Comparing to $\Lambda$ obtained by the unified EOSs (see Fig.~\ref{fig:5KL0}),
the differences in $\Lambda$ when using the nonunified EOSs are much smaller.
This is because $\Lambda$ depends on both the tidal Love number $k_2$ and
the compactness parameter $C=M/R$. Due to opposite $L$ dependence of the radius $R$
shown in Figs.~\ref{fig:4MR0} and~\ref{fig:8MR1},
the enhancement of $\Lambda$ with $L$ contributed by $k_2$ is counteracted by the decrease
of $R$ (equal to the increase of $C$) in the case of nonunified EOSs,
but it is enhanced by the increase of $R$ for unified EOSs.
Therefore, the $L$ dependence of $\Lambda$ shown in Fig.~\ref{fig:5KL0}
is more pronounced than the one in Fig.~\ref{fig:9KL1}.

\subsection{Effects of the core EOS}
\label{sec:3.3}

To examine the effect of the core EOS on neutron-star properties, we construct another set
of nonunified EOSs by matching the same crust EOS to different core segments.
Again, the crust-core transition is determined by the crossing point of the two segments.
In Fig.~\ref{fig:10PE2}, we display the pressure $P$ as a function of
the energy density $\varepsilon$ for the set of nonunified EOSs using the BPS EOS
for the outer crust, the TM1($L$=40) model for the inner crust,
and the TM1($L$=40, 60, 80, 111) models for the core.
It is shown that differences appear only in the core segments among these EOSs.
The model with a large $L$ predicts a stiff EOS at high densities.

In Fig.~\ref{fig:11MR2}, we plot the mass-radius relation obtained using the nonunified EOSs
with different core segments. It is seen that the impact of the slope parameter $L$ of the core
is rather obvious, especially on the radii of small mass neutron stars.
For the canonical $1.4 M_\odot$ neutron star, the radius $R_{1.4}$ is $\approx 14.53$ km in the case
of nonunified EOS with the TM1($L$=40) crust matching the TM1($L$=111) core, whereas it is reduced
to $\approx 13.12$ km when the TM1($L$=40) core is adopted. The difference between these two cases
is even larger than the one of unified EOSs shown in Fig.~\ref{fig:4MR0}.
This is because the combination of the TM1($L$=40) crust matching the TM1($L$=111) core
predicts the stiffest EOS among all combinations considered in this work.
This can be understood from the density dependence of
the symmetry energy, as shown in Fig.~\ref{fig:1ESYM}.
We find that both the core and crust EOSs can significantly affect the neutron-star radii,
as shown in Figs.~\ref{fig:8MR1} and~\ref{fig:11MR2}, but their $L$ dependences are opposite.

To study the impact of the core EOS on the tidal deformability of neutron stars,
we plot in Fig.~\ref{fig:12KL2} the tidal Love number $k_2$ (left panel)
and the dimensionless tidal deformability $\Lambda$ (right panel) as a function of
the neutron-star mass $M$, using the set of nonunified EOSs shown in Fig.~\ref{fig:10PE2}.
It is found that $k_2$ is insensitive to the slope parameter $L$ of the core,
which is different from the behavior shown in Figs.~\ref{fig:5KL0} and~\ref{fig:9KL1}.
This indicates that the tidal Love number $k_2$ is mainly determined by the crust EOS.
On the other hand, the tidal deformability $\Lambda$ shown in right panel of
Fig.~\ref{fig:12KL2} is clearly dependent on the slope parameter $L$ of the core.
The behavior of $\Lambda$ in this case is very similar to that using unified EOSs,
as shown in the right panel of Fig.~\ref{fig:5KL0}.
With increasing $L$ of the core, the enhancement of $\Lambda$ shown in Fig.~\ref{fig:12KL2}
is mostly contributed from the decrease of the compactness parameter $C$,
because $k_2$ is insensitive to the slope parameter $L$ of the core.
This is different from the case of unified EOSs, where the $L$ dependence of $\Lambda$
shown in Fig.~\ref{fig:5KL0} is determined by both $C$ and $k_2$.
To analyze the effects of the crust and core EOSs in more detail,
we present in Table~\ref{tab:3} some basic properties of neutron stars
obtained using different combinations of the crust and core segments.
It is found that $M_{\mathrm{max}}$ are determined by the core EOS,
whereas the properties of a canonical $1.4 M_\odot$ neutron star are affected
by both the crust and core EOSs. It is noticeable that the crust with different $L$
may result in $\approx 0.3$ km difference in the radius $R_{1.4}$ and
$\approx 0.2-0.4$ km difference in the crust thickness $\Delta R_{1.4}^{\mathrm{crust}}$.
Although $k_2^{1.4}$ and $C_{1.4}$ are affected by the crust EOS, the calculated
$\Lambda_{1.4}$ is not so sensitive to the crust EOS.
\begin{table*}[htbp]
\caption{Properties of neutron stars for different combinations of the crust and core EOSs.
$M_{\mathrm{max}}$ is the maximum mass of neutron stars.
$R_{1.4}$ and $\Delta R_{1.4}^{\mathrm{crust}}$ denote respectively the radius and crust thickness of
a canonical $1.4 M_\odot$ neutron star.
$k_2^{1.4}$, $C_{1.4}$, and $\Lambda_{1.4}$ are
the Love number, the compactness parameter, and the tidal deformability for a $1.4 M_\odot$
neutron star, respectively.}
\label{tab:3}
\begin{center}
\begin{tabular}{lccccccc}
\hline\hline
 EOS & Combination & $M_\mathrm{max}$ & $R_{1.4}$ & $\Delta R_{1.4}^{\mathrm{crust}}$
 & $k_2^{1.4}$ & $C_{1.4}$ & $\Lambda_{1.4}$
   \vspace{-0.10cm}\\
 TM1 & crust+core & $(M_\odot)$ & (km) & (km) &  &  & \\
\hline
unified         &  ($L$=40)+($L$=40)  & 2.12 & 13.12 & 1.25 & 0.095 & 0.158 & 652 \\
unified         & ($L$=111)+($L$=111) & 2.18 & 14.21 & 1.27 & 0.103 & 0.145 & 1047 \\
\hline
nonunified     &  ($L$=40)+($L$=111) & 2.18 & 14.53 & 1.44 & 0.092 & 0.142 & 1050 \\
nonunified     & ($L$=111)+($L$=40)  & 2.12 & 12.82 & 0.84 & 0.110 & 0.161 & 671 \\
\hline\hline
\end{tabular}
\end{center}
\end{table*}

\section{Conclusions}
\label{sec:4}

In this work, we constructed a set of unified EOSs based on RMF models
with different slope parameters $L$. We performed the self-consistent Thomas--Fermi
calculations for pasta phases appearing in the inner crust and then
determined the crust-core transition by comparing the energy densities between
pasta phases and homogeneous matter.
It was found that the model with a small $L$ predicts a large crust-core
transition density. By applying the set of unified EOSs in neutron-star calculations,
some correlations between the symmetry energy slope $L$ and neutron-star properties
were observed. It was found that a small $L$ corresponds to a small neutron-star radius
and therefore a small tidal deformability, which is favored by the recent analysis
of the GW170817 event.

To separately investigate the effects of crust and core EOSs on neutron-star
properties, we constructed two sets of nonunified EOSs: (1) the same core EOS
matching different crust EOSs; (2) the same crust EOS matching different core EOSs.
It was observed that different crust EOSs could lead to significant difference in neutron-star
radii. For the canonical $1.4 M_\odot$ neutron star, the radius $R_{1.4}$ changes from
$\approx 13.12$ km with the unified TM1($L$=40) EOS to $\approx 12.82$ km when replacing
the crust EOS with TM1($L$=111). Therefore, the uncertainty in $R_{1.4}$ induced
by different crust EOSs considered here is $\approx 0.3$ km.
On the other hand, the influence of the core EOS on neutron-star radii is more
pronounced than the one of the crust EOS. The uncertainty in $R_{1.4}$ induced
by different core EOSs is $\approx 1.4$ km.
We noticed that the $L$ dependence of neutron-star radii obtained using the two
sets of nonunified EOSs is opposite, which could be understood from the
density dependence of the symmetry energy.

We studied the tidal deformability of neutron stars using the two sets of
nonunified EOSs, in order to examine the effects of crust and core EOSs separately.
It was found that the effect of the core EOS on the tidal
deformability $\Lambda$ is more significant than the one of the crust EOS.
In fact, the tidal Love number $k_2$ is mainly determined by the crust EOS
and is insensitive to the slope parameter $L$ of the core.
With increasing $L$ of the core, the enhancement of $\Lambda$ is mostly
contributed from the increase of the neutron-star radius $R$.
On the other hand, the crust EOS could significantly affect both the
Love number $k_2$ and the radius $R$. However, the enhancement of $k_2$ with $L$
is largely counteracted by the decrease of $R$. Therefore, the resulting
tidal deformability $\Lambda$ is not so sensitive to the crust EOS.
We concluded that both the crust and core EOSs could significantly affect
neutron-star properties such as the radius and tidal deformability.
It is likely that the nuclear model with a small symmetry energy slope
is favored by various observational constraints.

\section*{Acknowledgment}

We would like to acknowledge helpful discussions with
K. Sumiyoshi, K. Nakazato, K. Oyamatsu, and H. Toki.
This work was supported in part by the National Natural Science Foundation of
China (Grants No. 11675083, No. 11775119, and No. 11805115).



\newpage

\begin{thebibliography}{99}

\bibitem{Abbo17} B. P. Abbott \textit{et al.} (LIGO
Scientifc Collaboration and Virgo Collaboration),
Phys. Rev. Lett. \textbf{119}, 161101 (2017).

\bibitem{Abbo18} B. P. Abbott \textit{et al.} (LIGO
Scientifc Collaboration and Virgo Collaboration),
Phys. Rev. Lett. \textbf{121}, 161101 (2018).

\bibitem{Demo10} P. B. Demorest, T. Pennucci, S. M. Ranson, M. S. E. Roberts,
and J. W. T. Hessels, Nature (London) \textbf{467}, 1081 (2010).

\bibitem{Fons16} E. Fonseca \textit{et al.}, Astrophys. J. \textbf{832}, 167 (2016).

\bibitem{Anto13} J. Antoniadis \textit{et al.}, Science \textbf{340}, 6131 (2013).

\bibitem{Crom19} H. T. Cromartie \textit{et al.}, arXiv:1904.06759.

\bibitem{Latt04} J. M. Lattimer and M. Prakash, Science \textbf{304}, 536 (2004).

\bibitem{Cham08} N. Chamel and P. Haensel, Living Rev. Relativ. \textbf{11}, 10 (2008).

\bibitem{Latt16} J. M. Lattimer and M. Prakash,
Phys. Rep. \textbf{621}, 127 (2016).

\bibitem{Oert17} M. Oertel, M. Hempel, T. Kl\"{a}hn, and S. Typel,
Rev. Mod. Phys. \textbf{89}, 015007 (2017).

\bibitem{Pear11} J. M. Pearson, S. Goriely, and N. Chamel,
Phys. Rev. C \textbf{83}, 065810 (2011).

\bibitem{Rave83} D. G. Ravenhall, C. J. Pethick, and J. R. Wilson,
Phys. Rev. Lett. \textbf{50}, 2066 (1983).

\bibitem{Mene08} S. S. Avancini, D. P. Menezes, M. D. Alloy, J. R. Marinelli,
M. M. W. Moraes, and C. Provid\^{e}ncia, Phys. Rev. C \textbf{78}, 015802 (2008).

\bibitem{Bao14a} S. S. Bao and H. Shen, Phys. Rev. C \textbf{89}, 045807 (2014).

\bibitem{Gril12} F. Grill, C. Provid\^{e}ncia, and S. S. Avancini,
Phys. Rev. C \textbf{85}, 055808 (2012).

\bibitem{Okma13} M. Okamoto, T. Maruyama, K. Yabana, and T. Tatsumi,
Phys. Rev. C \textbf{88}, 025801 (2013).

\bibitem{Bao15} S. S. Bao and H. Shen, Phys. Rev. C \textbf{91}, 015807 (2015).

\bibitem{Fatt17} F. J. Fattoyev, C. J. Horowitz, and B. Schuetrumpf,
Phys. Rev. C \textbf{95}, 055804 (2017).

\bibitem{Glen92} N. K. Glendenning, Phys. Rev. D \textbf{46}, 1274 (1992).

\bibitem{Glen01} N. K. Glendenning, Phys. Rep. \textbf{342}, 393 (2001).

\bibitem{Webe05} F. Weber,
Prog. Part. Nucl. Phys. \textbf{54}, 193 (2005).

\bibitem{Masu13} K. Masuda, T. Hatsuda, and T. Takatsuka,
Astrophys. J. \textbf{764}, 12 (2013);
Prog. Theor. Exp. Phys. \textbf{2013}, 073D01 (2013).

\bibitem{Bhat10} A. Bhattacharyya, I. N. Mishustin, and W. Greiner,
J. Phys. G \textbf{37}, 025201 (2010).

\bibitem{Yasu14} N. Yasutake, R. {\L}astowiecki, S. Beni{\'{c}}, D. Blaschke, T. Maruyama, and T. Tatsumi,
Phys. Rev. C \textbf{89}, 065803 (2014).

\bibitem{Webe16} W. M. Spinella, F. Weber, G. A. Contrera, and M. G. Orsaria,
Eur. Phys. J. A \textbf{52}, 61 (2016).

\bibitem{Wu19} X. H. Wu and H. Shen, Phys. Rev. C \textbf{99}, 065802 (2019).

\bibitem{Haen01} F. Douchin and P. Haensel,
Astron. Astrophys. \textbf{380}, 151 (2001).

\bibitem{Shen02} H. Shen, Phys. Rev. C \textbf{65}, 035802 (2002).

\bibitem{Miya13} T. Miyatsu, S. Yamamuro, and K. Nakazato,
Astrophys. J. \textbf{777}, 4 (2013).

\bibitem{Cham13} A. F. Fantina, N. Chamel, J. M. Pearson, and S. Goriely,
Astron. Astrophys. \textbf{559}, A128 (2013).

\bibitem{Shar15} B. K. Sharma, M. Centelles, X. Vi\~{n}as, M. Baldo, and G. F. Burgio,
Astron. Astrophys. \textbf{584}, A103 (2015).

\bibitem{Fort16} M. Fortin, C. Provid\^{e}ncia, Ad. R. Raduta, F. Gulminelli,
J. L. Zdunik, P. Haensel, and M. Bejger,
Phys. Rev. C \textbf{94}, 035804 (2016).

\bibitem{Sero86} B. D. Serot and J. D. Walecka,
Adv. Nucl. Phys. \textbf{16}, 1 (1986).

\bibitem{Meng06} J. Meng, H. Toki, S. G. Zhou, S. Q. Zhang, W. H. Long, and
L. S. Geng, Prog. Part. Nucl. Phys. \textbf{57}, 470 (2006).

\bibitem{TM1} Y. Sugahara and H. Toki, Nucl. Phys. A \textbf{579}, 557 (1994).

\bibitem{LiBA08} B. A. Li, L. W. Chen, and C. M. Ko,
Phys. Rep. \textbf{464}, 113 (2008).

\bibitem{Horo01} C. J. Horowitz and J. Piekarewicz,
Phys. Rev. Lett. \textbf{86}, 5647 (2001).

\bibitem{Duco10} C. Ducoin, J. Margueron, and C. Provid\^{e}ncia,
Europhys. Lett. \textbf{91}, 32001 (2010).

\bibitem{Oyam07} K. Oyamatsu and K. Iida,
Phys. Rev. C \textbf{75}, 015801 (2007).

\bibitem{Mene11} R. Cavagnoli, D. P. Menezes, and C. Provid\^{e}ncia,
Phys. Rev. C \textbf{84}, 065810 (2011).

\bibitem{Prov13} C. Provid\^{e}ncia and A. Rabhi,
Phys. Rev. C \textbf{87}, 055801 (2013).

\bibitem{Tews17}  I. Tews, J. M. Lattimer, A. Ohnishi, and E. E. Kolomeitsev,
Astrophys. J. \textbf{848}, 105 (2017).

\bibitem{Hebe13} K. Hebeler, J. M. Lattimer, C. J. Pethick, and A. Schwenk,
Astrophys. J. \textbf{773}, 11 (2013).

\bibitem{Latt14} J. M. Lattimer and A. W. Steiner,
Eur. Phys. J. A \textbf{50}, 40 (2014).

\bibitem{Hage15} G. Hagen, A. Ekstr\"{o}m, C. Forss\'{e}n, G. R. Jansen, W. Nazarewicz, T. Papenbrock, K. A. Wendt, S. Bacca,
N. Barnea, B. Carlsson, C. Drischler, K. Hebeler, M. Hjorth-Jensen, M. Miorelli, G. Orlandini, A. Schwenk, and J. Simonis,
Nat. Phys. \textbf{12}, 186 (2015)

\bibitem{Dani14} P. Danielewicz and J. Lee,
Nucl. Phys. A \textbf{922}, 1 (2014).

\bibitem{Roca15} X. Roca-Maza, X. Vi\~{n}as, M. Centelles, B. K. Agrawal, G. Col\`{o}, N. Paar, J. Piekarewicz, and D. Vretenar,
Phys. Rev. C \textbf{92}, 064304 (2015).

\bibitem{Birk17} J. Birkhan, M. Miorelli, S. Bacca, S. Bassauer, C. A. Bertulani, G. Hagen, H. Matsubara, P. von Neumann-Cosel,
T. Papenbrock, N. Pietralla, V. Yu. Ponomarev, A. Richter, A. Schwenk, and A. Tamii,
Phys. Rev. Lett. \textbf{118},  252501 (2017).

\bibitem{Dani17} P. Danielewicz, P. Singh, and J. Lee,
Nucl. Phys. A \textbf{958}, 147 (2017).

\bibitem{Bao14b} S. S. Bao, J. N. Hu, Z. W. Zhang, and H. Shen,
Phys. Rev. C \textbf{90}, 045802 (2014).

\bibitem{Shen11} H. Shen, H. Toki, K. Oyamatsu, and K. Sumiyoshi,
Astrophys. J. Suppl. \textbf{197}, 20 (2011).

\bibitem{IUFSU} F. J. Fattoyev, C. J. Horowitz, J. Piekarewicz, and G. Shen,
Phys. Rev. C \textbf{82}, 055803 (2010).

\bibitem{Gand12} S. Gandolfi, J. Carlson, and S. Reddy,
Phys. Rev. C \textbf{85}, 032801 (2012).

\bibitem{BPS71} G. Baym, C. Pethick, and P. Sutherland,
Astrophys. J. \textbf{170}, 299 (1971).

\bibitem{Zhang18} N. B. Zhang, B. A. Li, J. Xu,
Astrophys. J. \textbf{859}, 90 (2018).

\bibitem{Tews18} I. Tews, J. Margueron, and S. Reddy,
Phys. Rev. C \textbf{98}, 045804 (2018).

\bibitem{Zhu18} Z. Y. Zhu, E. P. Zhou, and A. Li,
Astrophys. J. \textbf{862}, 98 (2018).

\bibitem{Anna18} E. Annala, T. Gorda, A. Kurkela, and A. Vuorinen,
Phys. Rev. Lett. \textbf{120}, 172703 (2018).

\bibitem{Lim18} Y. Lim and J. W. Holt,
Phys. Rev. Lett. \textbf{121}, 062701 (2018).

\bibitem{Radi18} D. Radice, A. Perego, F. Zappa, and S. Bernuzzi,
Astrophys. J. Lett. \textbf{852}, L29 (2018).

\bibitem{De18} S. De, D. Finstad, J. M. Lattimer, D. A. Brown, E. Berger, and C. M. Biwer,
Phys. Rev. Lett. \textbf{121}, 091102 (2018).

\bibitem{Fatt18} F. J. Fattoyev, J. Piekarewicz, and C. J. Horowitz,
Phys. Rev. Lett. \textbf{120}, 172702 (2018).

\bibitem{Mali18} T. Malik, N. Alam, M. Fortin, C. Provid\^{e}ncia, B. K. Agrawal, T. K. Jha,
B. Kumar, and S. K. Patra, Phys. Rev. C \textbf{98}, 035804 (2018).

\bibitem{Dexh19} V. Dexheimer, R. O. Gomes, S. Schramm, and H. Pais,
J. Phys. G \textbf{46}, 034002 (2019).

\bibitem{LiBA19} P. G. Krastev and B. A. Li,
J. Phys. G \textbf{46}, 074001 (2019).

\bibitem{Hind08} T. Hinderer, Astrophys. J. \textbf{677}, 1216 (2008);
\textbf{697}, 964(E) (2009).

\bibitem{Post10} S. Postnikov, M. Prakash, and J. M. Lattimer,
Phys. Rev. D \textbf{82}, 024016 (2010).

\bibitem{Pais14} H. Pais, W. G. Newton, and J. R. Stone,
Phys. Rev. C \textbf{90}, 065802 (2014).

\bibitem{Latt91} J. M. Lattimer, C. J. Pethick, M. Prakash, P. Haensel,
Phys. Rev. Lett. \textbf{66}, 2701 (1991).

\bibitem{Negr18} R. Negreiros, L. Tolos, M. Centelles, A. Ramos, and V. Dexheimer,
Astrophys. J. \textbf{863}, 104 (2018).

\bibitem{Haen16} P. Haensel, M. Bejger, M. Fortin, and L. Zdunik,
Eur. Phys. J. A \textbf{52}, 59 (2016).

\end{thebibliography}
\end{document}